\documentclass[runningheads]{llncs}

\usepackage{eccv}

\usepackage{eccvabbrv}

\usepackage{graphicx}
\usepackage{booktabs}
\usepackage{colortbl}
\usepackage{multicol}
\usepackage{multirow}
\usepackage{arydshln}
\usepackage{amsmath}
\usepackage{pifont}
\usepackage{booktabs}
\usepackage{bm}
\usepackage{wrapfig}
\usepackage{algorithmic}
\usepackage[ruled,vlined,linesnumbered,commentsnumbered]{algorithm2e}

\usepackage[accsupp]{axessibility}  

\usepackage[pagebackref,breaklinks,colorlinks,citecolor=eccvblue]{hyperref}

\usepackage{orcidlink}

\usepackage{notoccite}

\newcommand{\method}{EQO}
\newcommand{\tabincell}[2]{\begin{tabular}{@{}#1@{}}#2\end{tabular}}

\begin{document}

\title{\method: Exploring Ultra-\underline{E}fficient Private Inference with Winograd-Based Protocol and \underline{Q}uantization Co-\underline{O}ptimization}

\titlerunning{\method: Exploring Ultra-Efficient Private Inference}

\author{Wenxuan Zeng \and
Tianshi Xu \and Meng Li\thanks{Corresponding author.} \and Runsheng Wang}

\authorrunning{W. Zeng et al.}

\institute{Peking University \\
\email{\{zwx.andy,tianshixu\}@stu.pku.edu.cn, \\ \{meng.li,r.wang\}@pku.edu.cn}
}

\maketitle

\begin{abstract}

Private convolutional neural network (CNN) inference based on secure two-party computation (2PC) suffers from high 
communication and latency overhead, especially from convolution layers.
In this paper, we propose \method, a quantized 2PC inference framework that jointly optimizes the CNNs and 2PC protocols.
\method~features a novel 2PC protocol that combines Winograd transformation with quantization for
efficient convolution computation.
However, we observe naively combining quantization and Winograd convolution is sub-optimal: Winograd
transformations introduce extensive local additions and weight outliers that increase the quantization bit widths and require frequent bit width conversions
with non-negligible communication overhead. Therefore, at the protocol level, we propose a series of optimizations for the
2PC inference graph to minimize the communication. At the network level, 
We develop a sensitivity-based mixed-precision quantization algorithm
to optimize network accuracy given communication constraints.
We further propose a 2PC-friendly bit re-weighting algorithm
to accommodate weight outliers without increasing bit widths.
With extensive experiments, \method~demonstrates 11.7$\times$, 3.6$\times$, and 6.3$\times$
communication reduction with 1.29\%, 1.16\%, and 1.29\% higher accuracy compared
to state-of-the-art frameworks SiRNN, COINN, and CoPriv, respectively.



\end{abstract}    
\section{Introduction}
\label{sec:intro}


\begin{figure}[!tb]
    \centering
    \includegraphics[width=\linewidth]{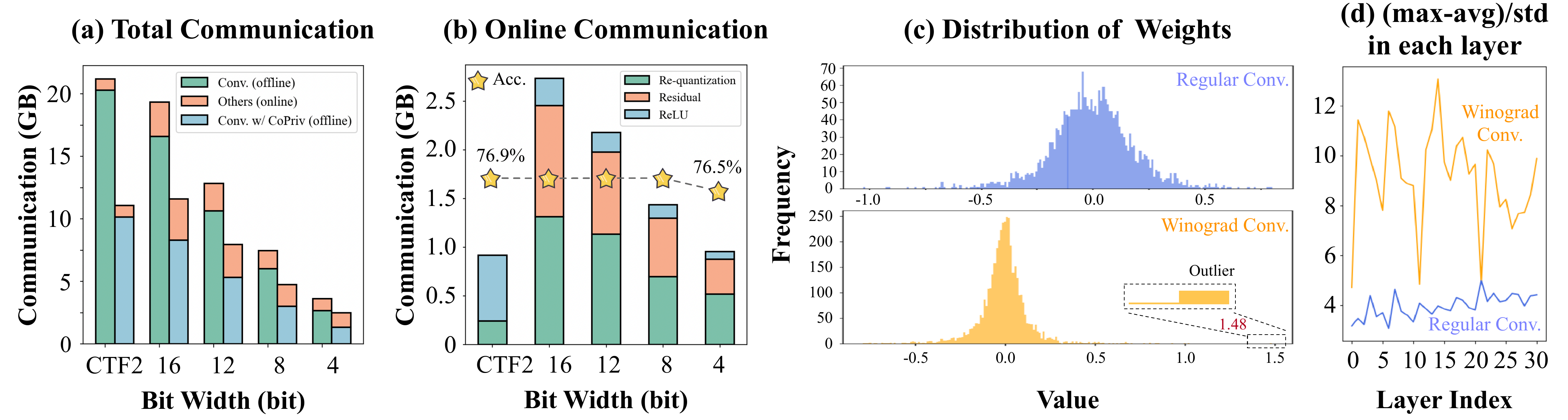}
    \caption{(a) Total communication and (b) online communication breakdown on the ResNet-50 building block profiled with
    CrypTFlow2 (CTF2) \cite{rathee2020cryptflow2} (uniform 37-bit) in the first column and SiRNN \cite{rathee2021sirnn} (supporting mixed precision)
    in the next four columns; 
    (c) weight distribution in regular and Winograd convolution, 
    (d) the ratio of (max-average) and standard deviation indicates weight outliers consistently exist after Winograd transformation across different layers.
    }
    \label{fig:intro_bar}
    \vspace{-15pt}
\end{figure}

Deep learning has recently demonstrated superior performance in various privacy-sensitive applications such as 
personal assistants \cite{kepuska2018next}, financial recommendation
system \cite{fan2023recommender,lee2022deep}, medical diagnosis \cite{richens2020improving,marques2020automated}, etc.
Privacy has thus emerged as one of the major concerns when deploying deep neural networks (DNNs).
Secure two-party computation (2PC) is proposed to provide cryptographically strong data privacy protection
and has attracted more and more attention in recent years \cite{rathee2020cryptflow2,rathee2021sirnn,demmler2015aby,mohassel2017secureml,mishra2020delphi,rathee2022secfloat,liu2017oblivious,shen2022abnn2}.


2PC inference frameworks target protecting the privacy of both model parameters held by the server and input data held by the client. 
By jointly executing a series of 2PC protocols, the client can learn the final inference results but nothing else on the model can be derived from the results.
Meanwhile, the server knows nothing about the client's input \cite{rathee2021sirnn,rathee2020cryptflow2,mohassel2017secureml,mishra2020delphi,jha2021deepreduce,lu2023squirrel,shen2022abnn2,mohassel2018aby3,liu2017oblivious}.

However, 2PC frameworks achieve high privacy protection at the cost of
orders of magnitude latency overhead. Due to the massive interaction
between the server and client, 2PC frameworks suffer from high
communication cost. As shown in Figure~\ref{fig:intro_bar}(a) and (b),
the total inference communication of a convolutional neural network (CNN)
is dominated by its convolution layers while the online 
communication is generated by non-linear functions, e.g., ReLU.
To improve communication efficiency, a series of works have proposed network and protocol optimizations
\cite{zeng2023copriv,riazi2019xonn,hussain2021coinn,rathee2021sirnn,
shen2022abnn2}. 
CoPriv \cite{zeng2023copriv} proposes a Winograd-based
convolution protocol to reduce the communication-expensive multiplications
at the cost of local additions. Although it achieves 2$\times$ communication
reduction, it still requires more than 2GB of communication for a single
ResNet-18 block. Recent works also leverage mixed-precision quantization for communication
reduction \cite{riazi2019xonn,hussain2021coinn,rathee2021sirnn,
shen2022abnn2}. Although total communication reduces
consistently with the inference bit widths,
as shown in Figure~\ref{fig:intro_bar}(b), existing mixed-precision protocols suffer from
much higher online communication even for 4-bit quantized networks
due to protocols like re-quantization, residual, etc.

As the communication of 2PC frameworks scales with both the inference bit widths
and the number of multiplications, we propose to combine the 
Winograd optimization with low-precision quantization. However, we
observe a naive combination leads to limited improvement. On one hand,
although the local additions in the Winograd transformations do not
introduce communication directly, they require higher inference bit width
and complicate the bit width conversion protocols. On the other hand, the
Winograd transformations also reshape the model weight distribution and
introduce more outliers that make quantization harder as shown in Figure \ref{fig:intro_bar}(c). As a result, a naive
combination only reduces $\sim$20\% total communication with even higher online 
communication compared to not using Winograd convolution as shown in Figure~\ref{fig:intro_bar}(a).

In this paper, we propose a communication-efficient 2PC framework named 
\method. \method~carefully combines Winograd transformation with
quantization and features a series of protocol and network
optimizations to address the aforementioned challenges. Our contributions
can be summarized below:
\begin{itemize}
    \item We observe the communication of 2PC inference scales with both
    the bit widths and the number of multiplications. Hence, we propose
    \method, which combines Winograd convolution and mixed-precision
    quantization for efficient 2PC inference for the first time. A series of graph-level optimizations
    are further proposed to reduce the online communication cost.
    \item We propose a communication-aware mixed-precision quantization algorithm, and further develop a 2PC-friendly bit re-weighting algorithm to
    handle the outliers introduced by the Winograd convolution.
    \item Extensive experiments demonstrate that \method~achieves 11.7$\times$, 3.6$\times$, and 6.3$\times$ communication reduction with 1.29\%, 1.16\%, and 1.29\% higher accuracy compared to state-of-the-art frameworks SiRNN, COINN, and CoPriv, respectively.
\end{itemize}

\section{Preliminaries}
\label{sec:pre}

Following \cite{rathee2021sirnn,rathee2020cryptflow2,mohassel2017secureml,mishra2020delphi,cho2022selective,jha2021deepreduce,kundu2023learning,shen2022abnn2,lou2020safenet},
\method~adopts a semi-honest attack model where both the server and client follow the protocol but also try to learn
more from the information than allowed.
We provide a detailed description of the threat model in Appendix \ref{supp:threat}.
For the convenience of readers, we summarize the underlying protocols used in this paper in Table \ref{tab:notation} and notations in Appendix \ref{supp:notation}.
Due to the space constraint, we put more preliminaries including Winograd convolution in Appendix \ref{supp:preli}.


\begin{table}[!tb]
    \centering
    \caption{Underlying protocols and the corresponding descriptions used in this paper. $\lambda$ denotes the security parameter.
    }
    \label{tab:notation}
    \resizebox{\linewidth}{!}{
    \begin{tabular}{c|c|c}
    \toprule
    Underlying Protocol  &  Description & Communication Complexity   \\
    \hline
    $\langle z \rangle^{(l_2)}=\Pi_{\mathrm{Ext}}^{l_1, l_2}(\langle x \rangle^{(l_1)})$  & \tabincell{c}{Extend $l_1$-bit $x$ to $l_2$-bit $z$ \\ such that $z^{(l_2)} = x^{(l_1)}$} &   $O(\lambda (l_1+1)+13l_1+l_2)$ \\
    $\langle z \rangle^{(l_1)}=\Pi_{\mathrm{Trunc}}^{l_1,l_2}(\langle x \rangle^{(l_1)})$ &  \tabincell{c}{Truncate (right shift) $l_1$-bit $x$ by $l_2$-bit \\ such that $z^{(l_1)}=x ^{(l_1)} \gg l_2$} & $O(\lambda (l_1+3)+15l_1+l_2+20)$ \\
    $\langle z \rangle^{(l_1-l_2)}=\Pi_{\mathrm{TR}}^{l_1,l_2}(\langle x \rangle^{(l_1)})$ & \tabincell{c}{Truncate $l_1$-bit $x$ by $l_2$-bit and discard the high $l_2$-bit \\ such that $z^{(l_1-l_2)}=x^{(l_1)} \gg l_2$} & $O(\lambda (l_2+1)+13l_2+l_1)$ \\
    $\langle z\rangle^{(l_2)}=\Pi_{\mathrm{Requant}}^{l_1,s_1,l_2,s_2}(\langle x \rangle^{(l_1)})$  &  \tabincell{c}{Re-quantize $x$ with $l_1$-bit and $s_1$ scale to \\ $z$ with $l_2$-bit and $s_2$ scale} & 
    \tabincell{c}{Combination of \\ $\Pi_{\mathrm{Ext}},\Pi_{\mathrm{Trunc}},\Pi_{\mathrm{TR}}$} \\
    \bottomrule
    \end{tabular}
    }
\end{table}

\subsection{Oblivious Transfer (OT)-based Linear Layers}
\label{sec:arss}

\begin{figure}[!tb]
    \centering
    \includegraphics[width=\linewidth]{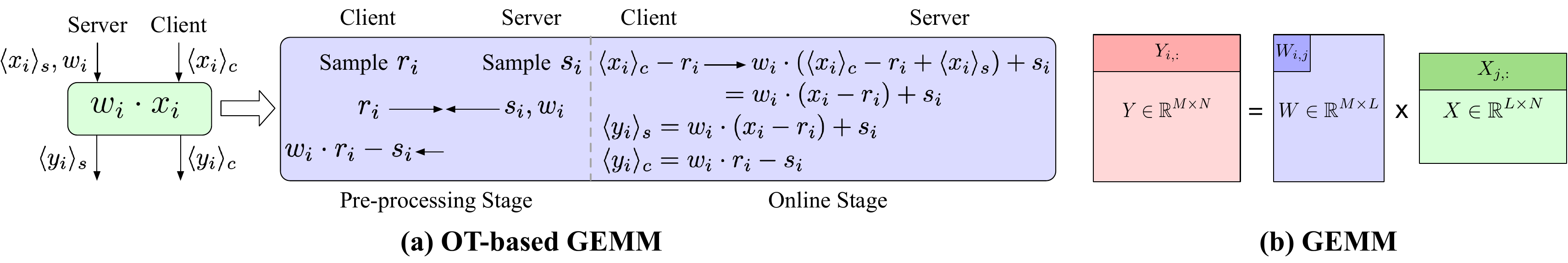}
    \caption{(a) Flow of OT-based linear layer, e.g., GEMM, including a pre-processing stage to generate input-independent helper data and an online stage to process client's input. (b) An illustration of a GEMM $Y=WX$.}
    \label{fig:ot_matmul}
\end{figure}

Figure~\ref{fig:ot_matmul}(a) shows the flow of 2PC-based inference. With arithmetic secret sharing (ArSS),
each intermediate activation tensor, e.g., $x_i$, is additively shared where the server holds $\langle x_i\rangle^S$
and the client holds $\langle x_i\rangle^C$ such that $x_i = \langle x_i\rangle^S+\langle x_i\rangle^C \mod P$ \cite{mohassel2017secureml}.
To generate the result $y_i$ of a linear layer, e.g., general matrix multiplication (GEMM),
a series of 2PC protocols is executed in the pre-processing stage and online stage \cite{mishra2020delphi}. 
In the pre-processing stage, the client and server first sample random $r_i$ and $s_i$, respectively.
Then, $\langle y_i\rangle_c=w_i\cdot r_i-s_i$ can be computed with a single OT if $r_i \in \{0, 1\}$.
With the vector optimization \cite{hussain2021coinn}, one OT can be extended to compute $w_i \cdot \bm{r_i} - \bm{s_i}$,
where $\bm{r_i}$ and $\bm{s_i}$ are both vectors.
When $w_i$ has $l_w$ bits, we can repeat the OT protocol $l_w$ times by computing $w_i^{(b)} \cdot r_i-s_i$ each time,
where $w_i^{(b)}$ denotes $b$-th bit of $w_i$. The final results can then be acquired by shifting and adding the partial results together.
Compared with the pre-processing stage, the online stage only requires very little communication to obtain $\langle y_i\rangle_s=w_i\cdot(x_i-r_i)+s_i$.

\subsection{Related Works}
Existing 2PC-based private inference frameworks can be divided into two categories, i.e., OT-based and homomorphic encryption (HE)-based.
The HE-based frameworks \cite{huang2022cheetah,hao2022iron,xu2023falcon,pang2023bolt} leverages HE to compute linear layers and 
achieve lower communication compared to OT-based frameworks at the cost of more computation overhead for both the server and client. 
Hence, HE-based and OT-based protocols have different applicable scenarios \cite{hussain2021coinn,zeng2023copriv}. 
For example, for resource-constrained clients, HE-based protocol may not be applicable since it involves repetitive
encoding, encryption, and decryption on the client side \cite{krieger2023aloha,gupta2022memfhe}.
Hence, we focus on optimizing OT-based frameworks to improve the communication efficiency in this work.

\begin{table}[!tb]
\centering
\caption{Comparison with existing works in terms of quantized optimization. ``\checkmark'' denotes the communication can be optimized. C/M means communication per multiplication (MUL). 
}
\label{tab:compare_exist}
\resizebox{\linewidth}{!}{
\begin{tabular}{cccccccc}
\toprule
\multirow{2}{*}{Framework} & \multicolumn{2}{c}{Protocol Optimization} & \multirow{2}{*}{Network Optimization} & \multicolumn{2}{c}{Linear} & \multirow{2}{*}{Non-linear}  \\
\cmidrule(l){2-3} \cmidrule{5-6}
                           & Operator Level    & Graph Level   &                               & \#MUL  & C/M  &                                   \\
\midrule                           
\cite{cho2022selective,kundu2023learning,jha2021deepreduce,jha2023deepreshape}  & -  & -   & ReLU-aware Pruning & -  & -   &   \checkmark   \\
\hline
SiRNN \cite{rathee2021sirnn}  &  16-bit Quant.  &  MSB Opt.  &  \tabincell{c}{-}   &  -    & \checkmark   &  \checkmark  \\
\hline
XONN \cite{riazi2019xonn}  &  Binary Quant.     &   -    &  \tabincell{c}{\tabincell{c}{End-to-end Binary Quant. \\ (Large Acc. Degradation)}}    &   -  &   \checkmark    &   \checkmark  \\
\hline
COINN \cite{hussain2021coinn}   & Factorized GEMM  & Protocol Conversion  &  \tabincell{c}{Mixed-precision Quant.}   & \checkmark  &  \checkmark   &  \checkmark   \\
\hline
ABNN2 \cite{shen2022abnn2} &  -  & -   & \tabincell{c}{Uniform Quant.}  & -  &  \checkmark  &   \checkmark    \\
\hline
QUOTIENT \cite{agrawal2019quotient}  &  -  &  -  &  \tabincell{c}{Gradient Quant. \\ Adaptive Gradient.}  &   -   &  \checkmark   &  \checkmark  \\
\hline
\rowcolor[rgb]{ .949,  .953,  .961}
\method~(ours)   &  \tabincell{c}{Mixed-precision Quant. \\ Winograd Conv.}   &   \tabincell{c}{Simplified Residual \\ Protocol Fusion \\ MSB Opt.}    &  \tabincell{c}{Comm.-aware Quant. \\ Mixed-precision Quant. \\ High-precision Residual \\ Bit Re-weighting}   &   \checkmark   &  \checkmark   &   \checkmark   \\
\bottomrule                           
\end{tabular}
}
\end{table}

In recent years, there has been an increasing amount of literature on efficient OT-based private inference, including protocol optimization \cite{demmler2015aby,mohassel2018aby3,rathee2020cryptflow2,rathee2021sirnn,knott2021crypten,rathee2022secfloat,mohassel2017secureml}, network optimization \cite{cho2022selective,jha2021deepreduce,kundu2023learning,li2022mpcformer,zeng2023mpcvit,lou2020safenet}, and network/protocol joint optimization \cite{mishra2020delphi,hussain2021coinn,zeng2023copriv}.
To reduce the communication overhead, quantization has been used for private inference
\cite{riazi2019xonn,samragh2021application,hussain2021coinn,rathee2021sirnn,shen2022abnn2,hasler2023overdrive,agrawal2019quotient}.
In Table \ref{tab:compare_exist}, we compare \method~with previous works
qualitatively. As can be observed, \method~jointly optimizes both protocol and 
network and simultaneously reduces the number of multiplication and communication per multiplication.
We leave the detailed review of existing works to Appendix \ref{related}.

\section{Motivations and Overview}
\label{sec:motivation}




In this section, we analyze the communication complexity of OT-based 2PC inference.
We also explain the challenges when combining Winograd transformation with quantization,
which motivates \method.

\paragraph{Observation 1: the total communication of OT-based 2PC is determined by both the bit widths and the number of multiplications in linear layers.}
Consider an example of $Y = WX$, 
where $W\in \mathbb{R}^{M\times L}$, $X\in \mathbb{R}^{L\times N}$ and $Y\in \mathbb{R}^{M\times N}$ in Figure \ref{fig:ot_matmul}(b).
With one round of OT, we can compute $W_{i, j}^{(b)} \cdot X_{j, :}$ for the $b$-th bit of $W_{i, j}$ and $j$-th row of $X$.
Then, the $i$-th row of $Y$, denoted as $Y_{i,:}$, can be computed as
$$
    \label{eq:ot}
    Y_{i, :} = \sum_{j=0}^{L-1} \sum_{b=0}^{l_w-1} 2^b \cdot W_{i, j}^{(b)} \cdot X_{j, :},
$$
where $l_w$ denotes the bit width of $W$.
Hence, to compute $Y_{i, :}$, in total $O(l_wL)$ OTs are invoked.
In each OT, the communication scales proportionally
with the vector length of $X_{j, :}$, i.e., $O(N l_x)$, where $l_x$ denotes the bit width of $X$.
The total communication of the GEMM thus becomes $O(MLN l_w l_x)$.
Thus, we observe the total communication of a GEMM scales with both the operands' bit widths,
i.e., $l_x$ and $l_w$, and the number of multiplications, i.e., $MLN$,
both of which impact the round of OTs and the communication per OT.
Convolutions follow a similar relation as GEMM. Hence, combining Winograd transformations and quantization
is a promising solution to improve the communication efficiency of convolution layers.
For non-linear layers, e.g., ReLU,
the communication cost also scales proportionally with activation bit widths \cite{rathee2021sirnn}.

\begin{figure}[!tb]
    \centering
    \begin{minipage}[t]{0.5\textwidth}
    \centering
    \captionof{figure}{Insertion of bit extension ($\textcolor{Goldenrod}{\bigstar}$) with quantized inference. There is a re-quantization in the GEMM protocol. 
    }
    \label{fig:compare_extend}
    \includegraphics[width=0.85\linewidth]{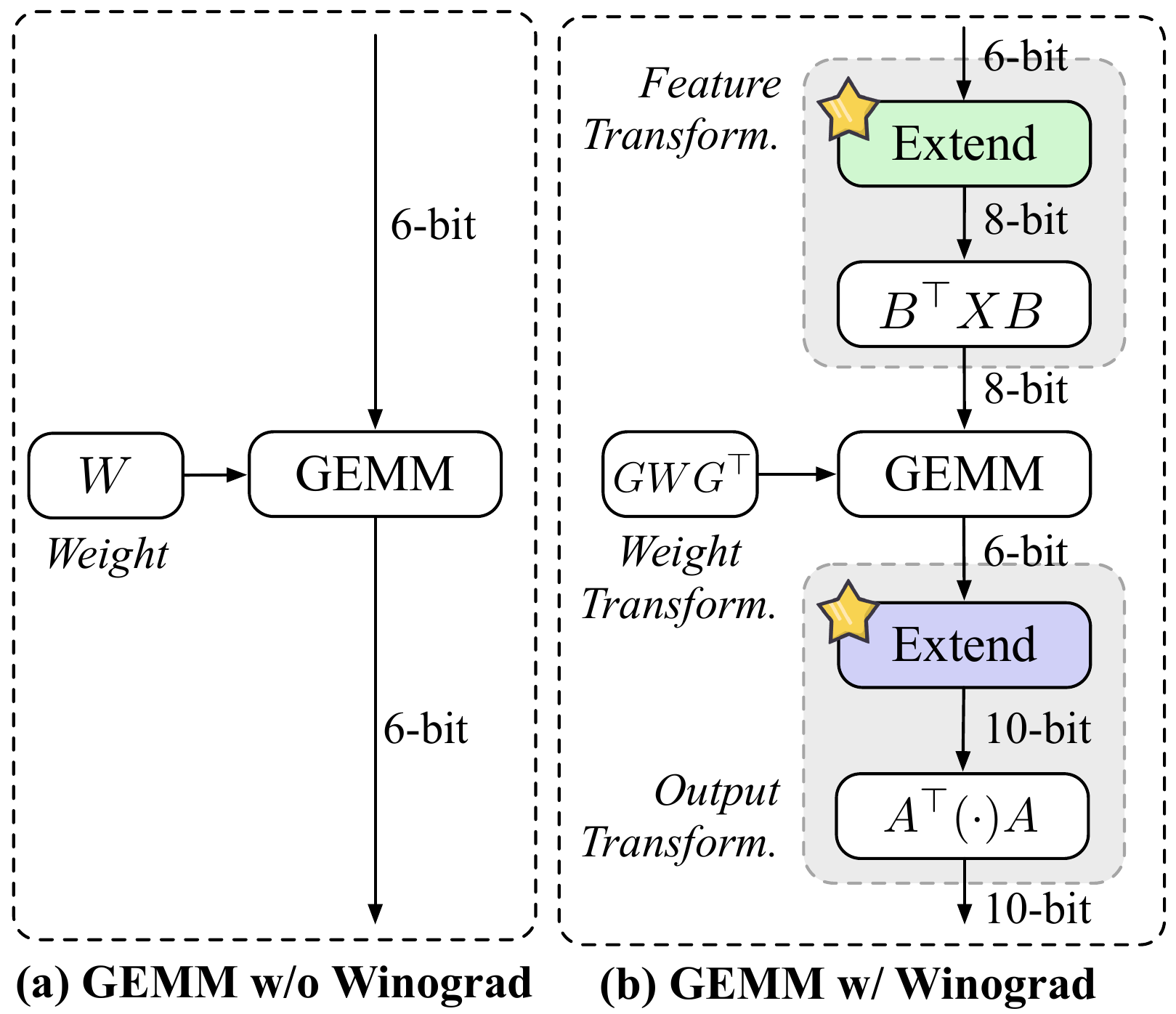}
    \end{minipage}
    \hfill
    \begin{minipage}[t]{0.45\textwidth}
    \centering
    \caption{Communication breakdown of Winograd convolution on ResNet-32. After graph-level protocol fusion introduced in Section \ref{sec:fusion}, the communication of both feature and output transformation are reduced. 
    }
    \label{fig:winograd_breakdown}
    \includegraphics[width=\linewidth]{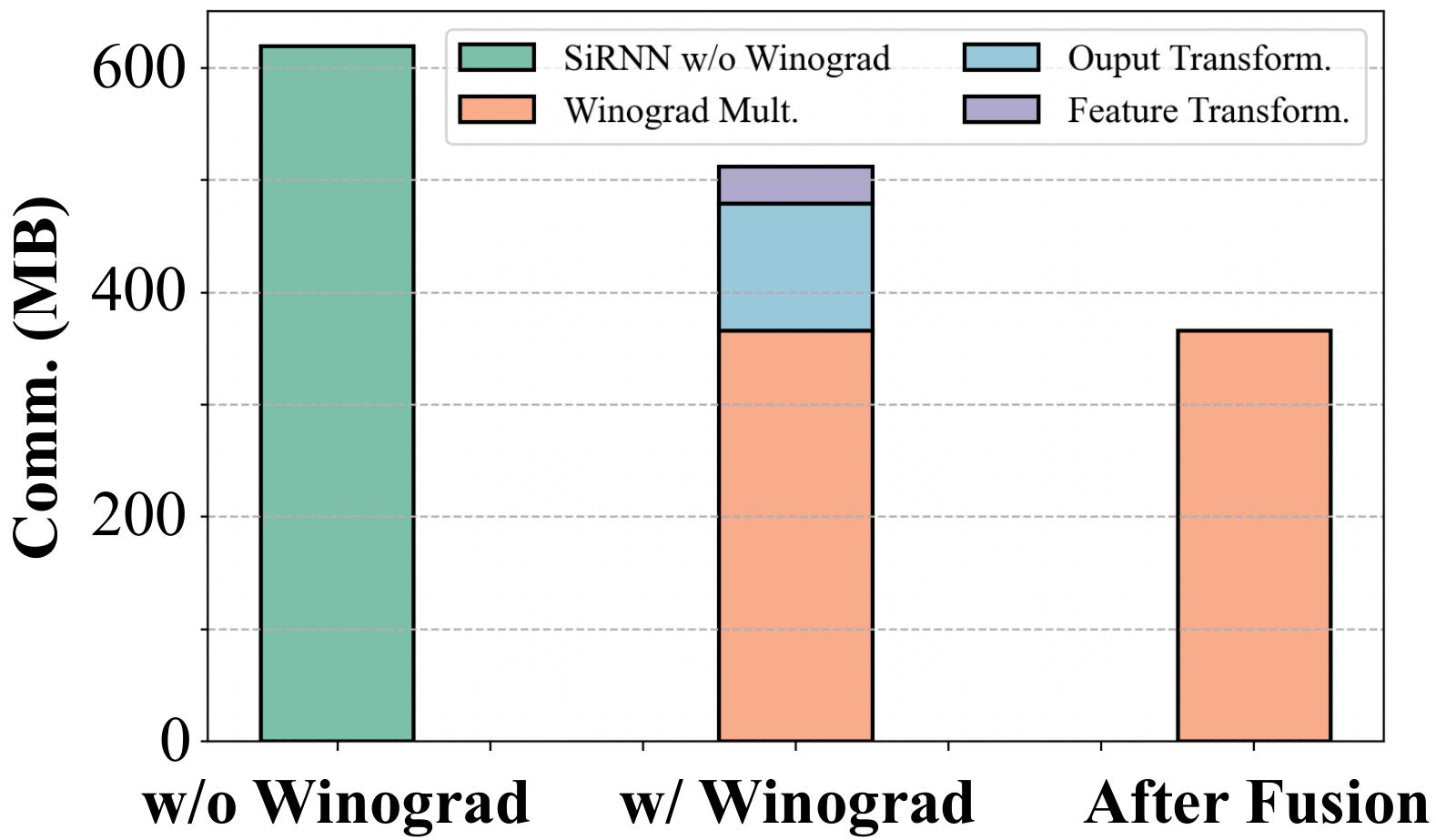}
    \end{minipage}
\end{figure}

\paragraph{Observation 2: a naive combination of Winograd transformations and quantization provides limited communication reduction.}
Although Winograd transformation reduces the number of multiplications, it is not friendly to quantization
as it introduces many more local additions during the feature and output transformation.
Hence, to guarantee the computation correctness with avoiding overflow, extra bit width conversion protocols need to be performed
as shown in Figure~\ref{fig:compare_extend}. Take ResNet-32 with 2-bit weight and 6-bit activation (abbreviated as W2A6) as an example in Figure \ref{fig:winograd_breakdown},
naively combining Winograd transformation and quantization achieves only $\sim$20\% communication reduction compared to not using Winograd convolution in SiRNN,
which is far less compared to \cite{zeng2023copriv}.
Hence, protocol optimization is important to reduce the overhead induced by the bit width conversions.

\begin{wrapfigure}{r}{0.3\textwidth}
    \centering
    \includegraphics[width=\linewidth]{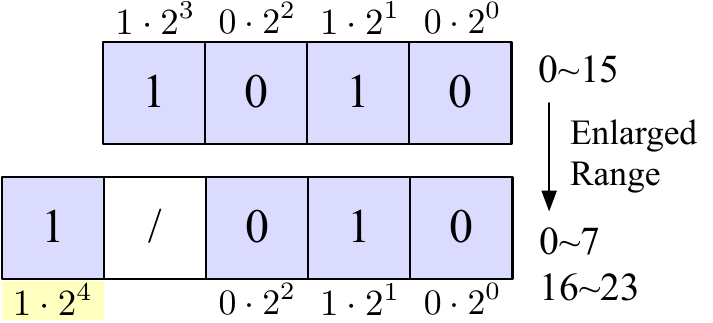}
    \caption{Example of bit re-weighting with adjusted representation range. 
    }
    \label{fig:skip}
\end{wrapfigure}

\paragraph{Observation 3: Winograd transformations introduce quantization outliers 
and make low-precision weight quantization challenging.}
We show the weight distribution with and without Winograd transformations in Figure~\ref{fig:intro_bar}(c).
The weight transformation involves multiplying or dividing large plaintext integers \cite{lavin2016fast} and tends to
generate large weight outliers, which makes the low-precision quantization challenging.
Instead of simply increasing the quantization bit width, we observe it is possible to accommodate
the weight outliers by bit re-weighting. Recall for OT-based linear layers, 
each weight is first written as $\sum_{b=0}^{l_w-1} w^{(b)} \cdot 2^b$ (we ignore the sign bit for convenience)
and then, each bit $w^{(b)}$
is multiplied with the corresponding activations with a single OT. The final results are acquired
by combining the partial results by shift and addition.
This provides us with unique opportunities to re-weight each bit by adjusting $2^b$ to increase
the representation range without causing extra communication overhead. 
An example is shown in Figure~\ref{fig:skip}. As can be observed, through bit re-weighting, we can increase
the representation range by $2\times$ with the same quantization bit width (note the total number of
possible represented values remain the same).

\begin{figure}[!tb]
    \centering    
    \includegraphics[width=\linewidth]{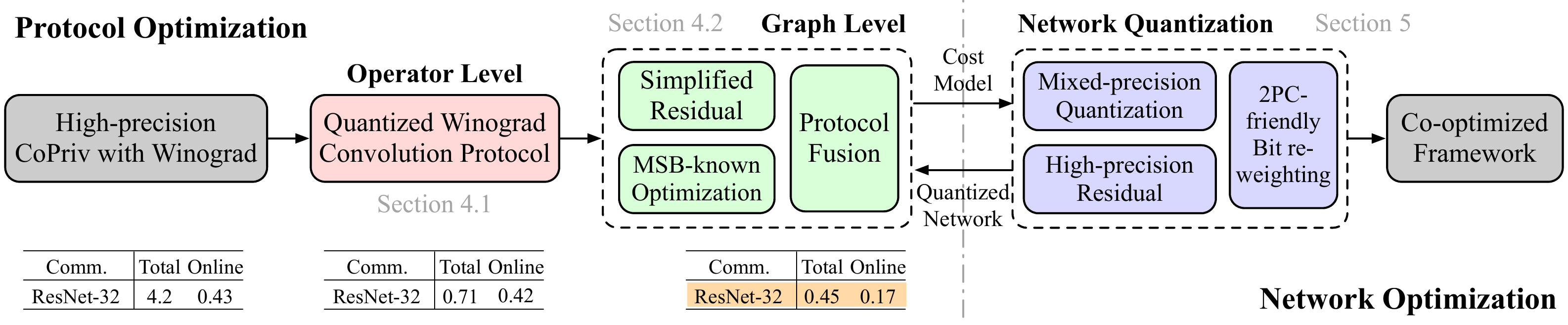}
    \caption{Overview of co-optimization framework \method~and the communication cost (GB) after each optimization step. Example is evaluated on ResNet-32 with W2A6.
    }
    \label{fig:overview}
\end{figure}

\paragraph{Overview of \method}
\label{subsec:overview}
Based on these observations, we propose \method, a network/protocol co-optimization framework for communication-efficient 2PC inference.
We show the overview of in Figure \ref{fig:overview}.
We first optimize the 2PC protocol for convolutions combining quantization and Winograd transformation (Section \ref{sec:winograd}).
We then propose a series of graph-level optimizations,
including protocol fusion to reduce the communication for Winograd transformations (Section \ref{sec:fusion}),
simplified residual protocol to remove redundant bit width conversion protocols
(Section \ref{sec:fusion}), and graph-level activation sign propagation and
protocol optimization given known most significant bits (MSBs)
(Section \ref{sec:msb}). 
In Figure \ref{fig:overview}, although Winograd is quantized to low precision, there is no benefit for online communication due to the extra bit width conversions.
As a result, the graph-level optimizations enable to reduce
the online communication by 2.5$\times$ in the example.
At the network level, we further propose communication-aware mixed-precision quantization and bit re-weighting algorithms to reduce the inference bit widths as well as communication (Section \ref{sec:quant}).
\method~reduces the total communication by 9$\times$ in the example.

\section{\method~Protocol Optimization}
\label{sec:method}


\subsection{Quantized Winograd Convolution Protocol}
\label{sec:winograd}




As explained in Section~\ref{sec:motivation}, when combining the Winograd transformation and quantization,
extra bit width conversions, i.e., extensions are needed to guarantee compute correctness due to local additions in the feature and
output transformations. Two natural questions hence arise: 
\underline{1)} where to insert the bit width extension protocols? \underline{2)} how many bits to extend?

\begin{wrapfigure}{r}{0.4\textwidth}
    \vspace{-20pt}
    \centering
    \includegraphics[width=\linewidth]{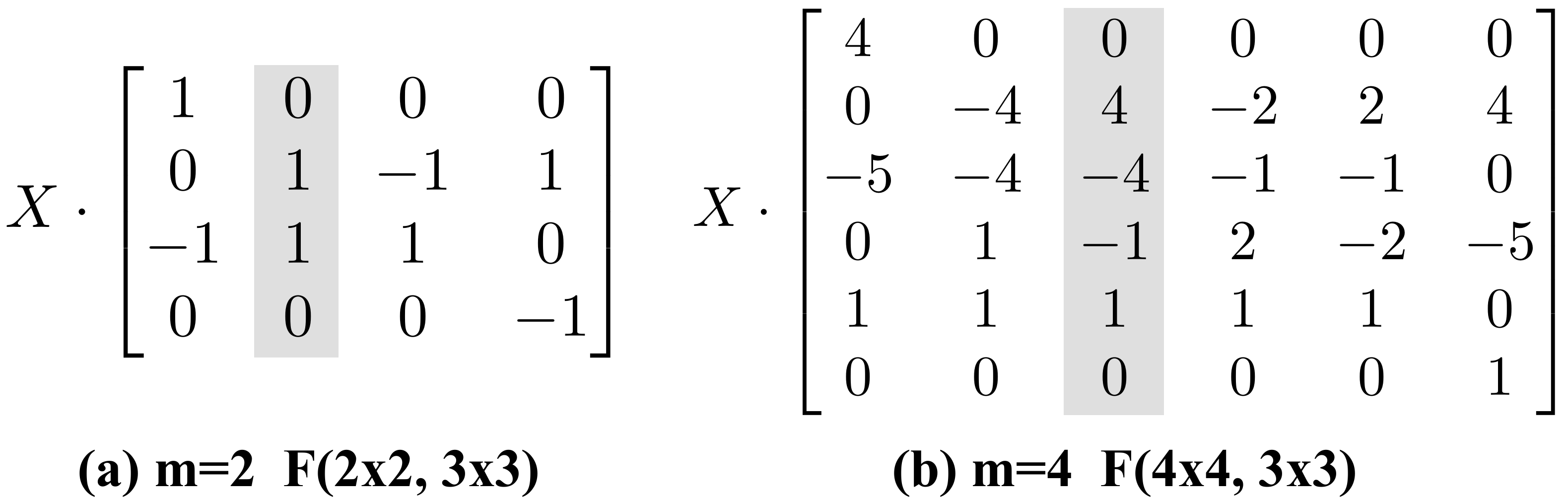}
    \caption{GEMM of $X$ and transformation matrix $B$. The columns colored in grey mark the possible maximum number of additions.
    }
    \label{fig:matmul_add}
    \vspace{-20pt}
\end{wrapfigure}

For the first question, there are different ways to insert the bit extension. 
Take the output transformation as an example. 
One method is to extend the activations right before computing the output transformation,
which incurs online communication.
The second method is to insert the bit width extensions before the GEMM protocol.
While this enables to merge the bit extension protocols with the offline weight transformation,
we find it will drastically increase the GEMM communication and thus, choose the first method in \method.

For the second question, we have the following lemma that bounds the outputs after transformation.
\begin{lemma}
    Consider $Y = XB$. For each element $Y_{i, j}$, its magnitude can always be bounded by
    \begin{align*}
    |Y_{i, j}| = |\sum_{k} X_{i, k} B_{k, j}| & \leq \sum_{k} | X_{i, k} | |B_{k, j}|  
                                  \leq 2^{l_x} \sum_k |B_{k, j}| = 2^{l_x + \log_2 ||B_{:, j}||_1},
    \end{align*}
    where $||\cdot||_1$ is the $\ell_1$-norm.
\end{lemma}
Therefore, each output element $Y_{i, j}$ requires at most $l_x + \log_2 ||B_{:, j}||_1$ bits.
Since we use per-tensor activation quantization, to guarantee the computation correctness, we need
$\max_{j} l_x + \log_2 ||B_{:, j}||_1$ bits to represent the output. Similarly, to compute $B^\top X$,
$\max_{j} \log_2 ||B_{:, j}||_1$ is needed, adding up to $2 \times \max_{j} \log_2 ||B_{:, j}||_1$ bits
to extend. In Figure \ref{fig:matmul_add}, we show the transformation matrix $B$ for
Winograd convolution with the output tile size of $2$ and $4$. As can be calculated,
2-bit and 8-bit extensions are needed, respectively. 
As 8-bit extension drastically increases the communication
of the following GEMM, we choose 2 as the output tile size. The bit extension for the output
transformation can be calculated similarly.

\begin{figure}[!tb]
    \centering
    \includegraphics[width=\linewidth]{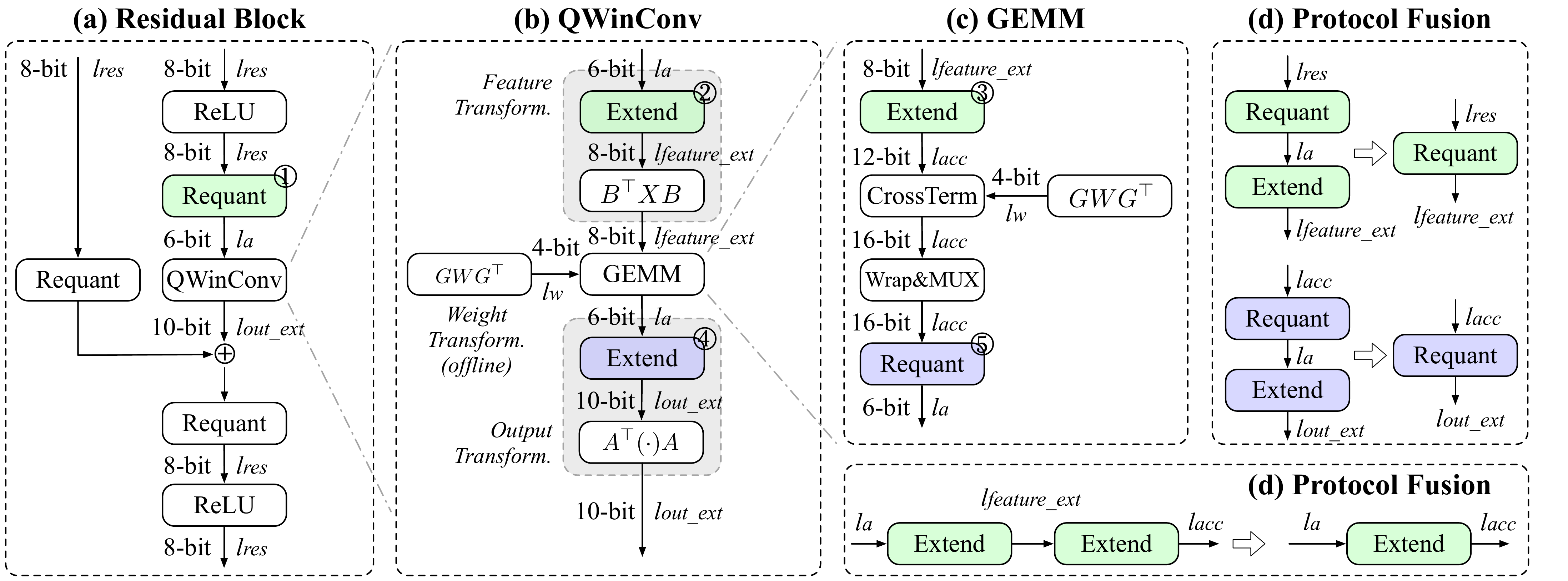}
    \caption{Overall framework of quantized Winograd convolution protocol \method~in (a) an example residual block; (b) the design of QWinConv; (c) GEMM protocol in SiRNN \cite{rathee2021sirnn};
    (d) graph-level protocol fusion.
    }
    \label{fig:protocol}
\end{figure}

Based on the above analysis, we propose the quantized Winograd convolution protocol, dubbed QWinConv, in Figure \ref{fig:protocol}.
Consider an example of 4-bit weights and 6-bit activations.
In Figure \ref{fig:protocol}(a), before QWinConv, following SiRNN \cite{rathee2021sirnn},
activations are first re-quantized to 6-bit to reduce the communication of QWinConv (block \ding{172}).
We always keep the residual in higher bit width, i.e., 8-bit, for better accuracy \cite{wu2018mixed,yang2021fracbits}.
After QWinConv, the residual and QWinConv output are added together with the residual protocol.
The design of QWinConv is shown in Figure \ref{fig:protocol}(b), which involves four steps:
feature transformation, weight transformation, Winograd-domain GEMM, and output transformation.
The GEMM protocol in Figure \ref{fig:protocol}(c) follows SiRNN \cite{rathee2021sirnn}
and the extension (block \ding{174}) ensures accumulation correctness.
Weight transformation and its quantization can be conducted offline,
while feature and output transformation and quantization must be executed online.
We insert bit extensions right before the feature and output transformation as marked with block \ding{173} and \ding{175}, respectively.

\subsection{Graph-level Protocol Optimization}


\begin{figure}[!tb]
    \centering
    \begin{minipage}[t]{0.6\textwidth}
    \includegraphics[width=\linewidth]{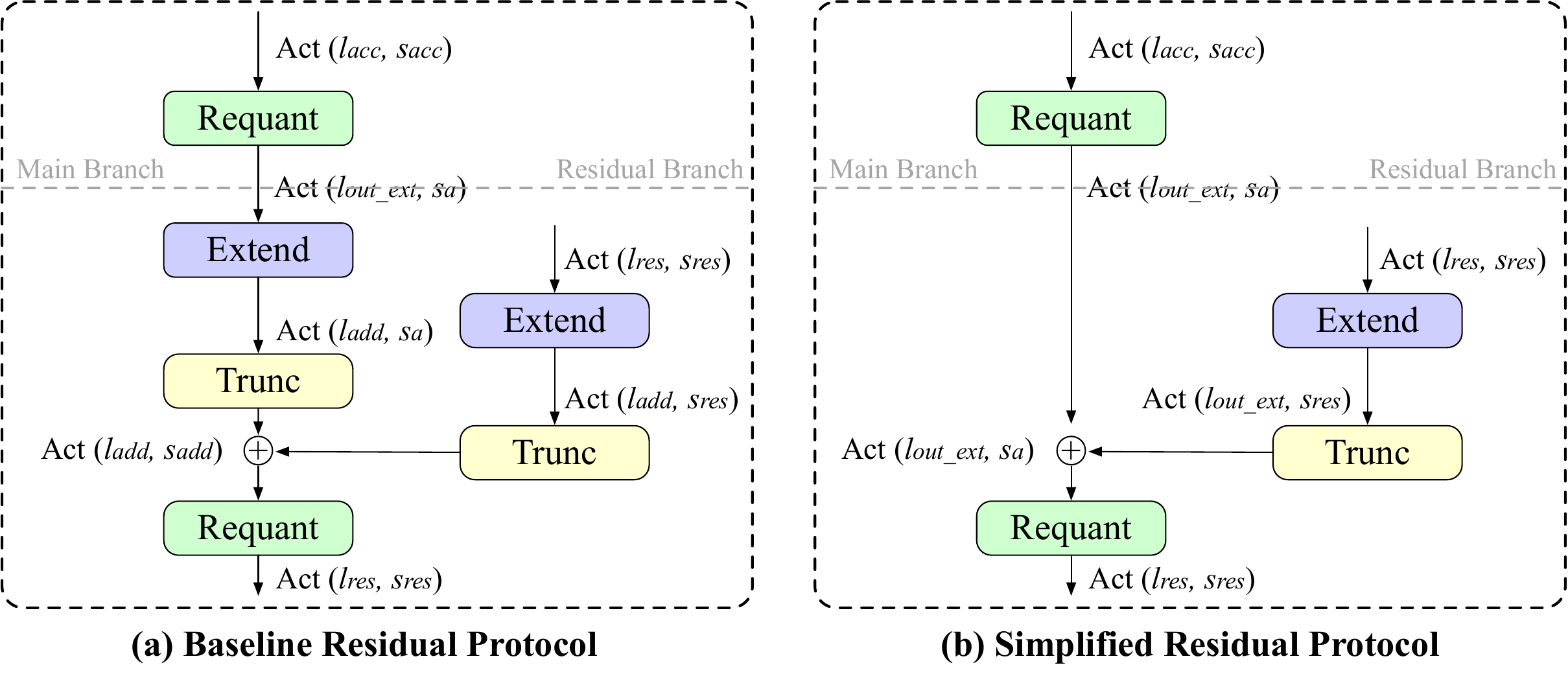}
    \caption{Comparison between (a) baseline residual protocol \cite{rathee2021sirnn} and (b) our simplified residual protocol. $l, s$ mean the bit width and scale of the operand, respectively. 
    }
    \label{fig:residual_protocol}
    \end{minipage}
    \hfill
    \begin{minipage}[t]{0.35\textwidth}
    \centering
    \includegraphics[width=0.8\linewidth]{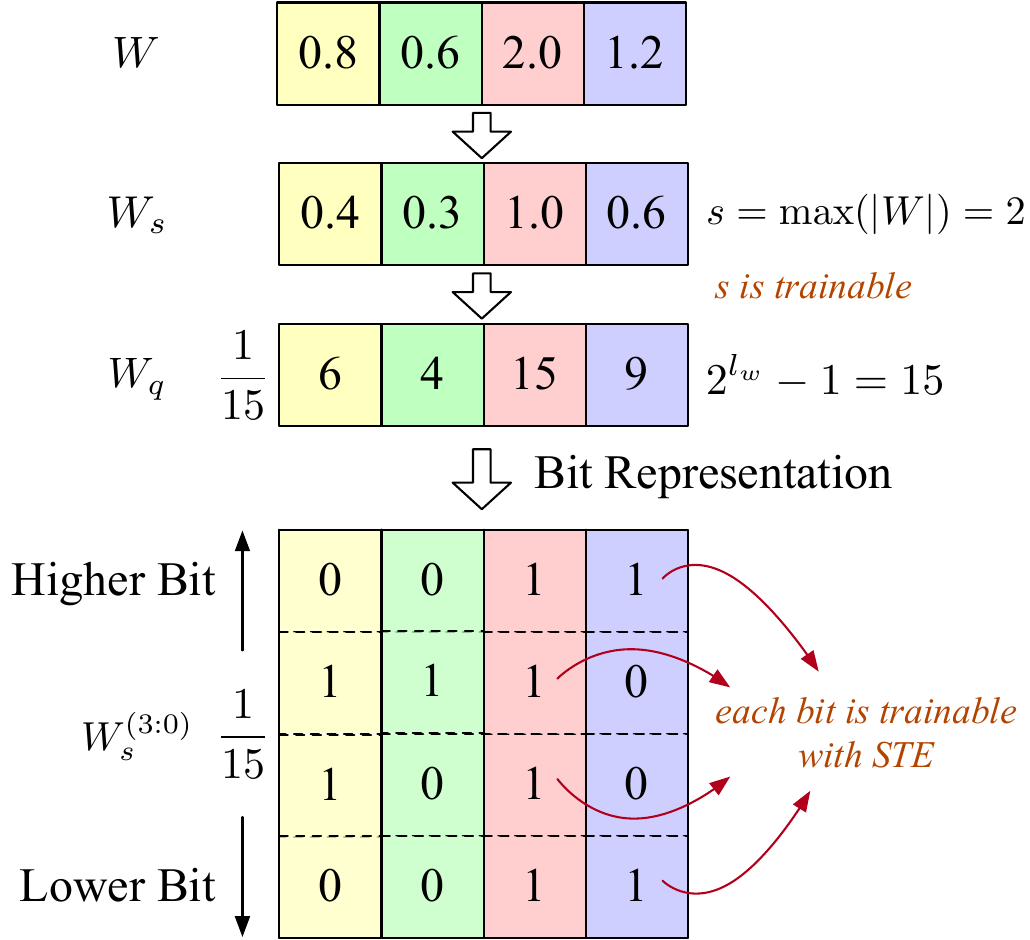}
    \caption{Conversion of bit representation with bit width $l_w=4$. 
    \label{fig:bit_repre}
    During training, each bit in $W_s^{(3:0)}$ is trainable.
    }
    \end{minipage}
\end{figure}

\paragraph{Graph-level Protocol Fusion}
\label{sec:fusion}
As explained in Section~\ref{sec:motivation}, extra bit width conversion protocols in QWinConv
(\ding{173} and \ding{175} in Figure~\ref{fig:protocol}) increase the online
communication and diminish the benefits of combining quantization with Winograd transformation,
especially for low-precision quantized networks, e.g., 4-bit. 
We observe there are neighboring bit width conversion protocols that can be fused to reduce the overhead.
Specifically, we find two patterns that appear frequently:
1) consecutive bit width conversions, e.g., \ding{172} and \ding{173}, \ding{175} and \ding{176};
2) bit width conversions that are only separated by local operations, e.g., \ding{173} and \ding{174}.
As the cost of bit width conversion protocols is only determined to the bit width of inputs \cite{rathee2021sirnn},
such protocol fusion enables complete removal of the cost of \ding{173}, \ding{174} and \ding{176},
which completely removes the cost of feature and output transformations.
We refer interested readers to a formal description in Appendix \ref{supp:fusion}.

\paragraph{Simplified Residual Protocol}
\label{sec:residual}
As shown in Figure \ref{fig:intro_bar}(b), residual protocol consumes around 50\% of the online communication due to the alignment 
of both bit widths and scales\footnote{We explain the details of bit width and scales in quantization in Appendix \ref{supp:quant}.} as shown in Figure \ref{fig:residual_protocol}(a).
Therefore, we propose a new simplified residual protocol to reduce the communication as shown in Figure \ref{fig:residual_protocol}(b).
Specifically, we directly align the bit width and scale of residual to the output of QWinConv for addition.
In this way, we get rid of the redundant bit width conversion protocol in the main branch, reducing the communication from $O(\lambda(l_{out\_ext}+l_{res}+2l_{add}+8))$ to $O(\lambda(l_{out\_ext}+l_{res}+4))$.


\paragraph{MSB-known Optimization}
\label{sec:msb}
As pointed out by \cite{rathee2021sirnn,huang2022cheetah,hao2022iron}, 2PC protocols can be designed in a more efficient way when the MSB of the secret shares is known.
Since the activation after ReLU is always non-negative, protocols including truncation and extension can be further optimized.
In \method, we locate the ReLU function and then, propagate the sign of the activations to all the downstream protocols.
In Figure \ref{fig:protocol}, for example, the input of re-quantization and extensions in green (block \ding{172}, \ding{173} and \ding{174}) must be non-negative,
so they can be optimized. 
In contrast, re-quantization and extension in blue (block \ding{175} and \ding{176}) can not be optimized since the GEMM outputs can be either positive or negative.

\paragraph{Security and Correctness Analysis}
\method~is built on top of SiRNN \cite{rathee2021sirnn} with new quantized Winograd convolution protocol and
graph-level protocol optimizations. The security of the quantized Winograd convolution protocol directly follows
from the security guarantee from OT. Observe that all the communication between the client and the server is performed
through OT which ensures the privacy of both the selection bits and messages. The graph-level protocol optimizations,
including protocol fusion, residual simplification, and MSB-known optimization leverage information known to both parties,
e.g., the network architecture, and thus, do not reveal extra information.
The correctness of the quantized Winograd convolution protocol is guaranteed by the theorem of Winograd transformation
\cite{lavin2016fast} and bit width extensions to avoid overflow.



\section{\method~Network Quantization Optimization}
\label{sec:quant}

In this section, we propose Winograd-domain quantization algorithm that is compatible with OT-based 2PC inference.
The overall training procedure is shown in Algorithm \ref{alg:quant}. 
We first assign different bit widths to each layer based on the quantization sensitivity, and then propose 2PC-friendly bit re-weighting to improve the quantization performance.

\subsection{Communication-aware Sensitivity-based Quantization}
\label{sec:qat}

Different layers in a CNN have different sensitivity to the quantization noise \cite{dong2019hawq,dong2020hawq,yao2021hawq}.
To enable accurate and efficient private inference of \method, 
we propose a communication-aware mixed-precision algorithm to search for the optimal bit width assignment for each layer based on sensitivity.
Sensitivity can be approximately measured by the average trace of the Hessian matrix.
Following \cite{dong2020hawq}, 
let $\Omega_i$ denote the output perturbation induced by quantizing the $i$-th layer. Then, we have
$$
 \Omega_i = \overline{Tr}(H_i)\cdot ||\mathrm{Quant}(GW_iG^\top)-GW_iG^\top||_2^2,
$$
where $H_i$ and $\overline{Tr}(H_i)$ denote the Hessian matrix of the $i$-th layer and its average trace,
$GW_iG^\top$ denotes the Winograd-domain weight, $\mathrm{Quant}(\cdot)$ denotes quantization,
and $||\cdot||_2$ denotes the $\ell_2$-norm of quantization perturbation.
Given the communication bound $\zeta$ and the set of admissible bit widths $\mathcal B$,
inspired by \cite{yao2021hawq},
we formulate the communication-aware bit width assignment problem as an integer linear programming problem (ILP):
\begin{align*}
    \min \sum_{i\in [1,L]} \sum_{j \in \mathcal B}  T_{i, j} \cdot \Omega_{i, j}, 
    \quad 
    \text{s.t.} \sum_{i\in [1,L]} \sum_{j \in \mathcal B} T_{i, j} \cdot C_{i, j} \leq \zeta, 
    \sum_{j \in \mathcal B} T_{i, j} =1, \forall i \in [1,L],
\end{align*}
where $T_{i, j} \in \{0, 1\}$, $C_{i, j}$, and $\Omega_{i, j}$ are the indicator, communication cost, and perturbation when
quantizing the $i$-th layer to $j$-bit, respectively.
The objective is to minimize the perturbation in the network output under the given communication constraint.

\begin{algorithm}[!tb]
\caption{Training Procedure of \method~Quantization}
\label{alg:quant}
\SetKwInOut{Input}{Input}
\SetKwInOut{Output}{Output}

\SetKwFunction{Tuple}{}
\SetKwComment{Comment}{/* }{ */}

\footnotesize
\Input{Pre-trained floating-point weight $w_f$, communication bound $\zeta$, finetuning epochs $E$}
\Output{Set of bit weights $\mathbb{B}$ and quantized weight $w$}
\BlankLine



$l_w = \texttt{HessianAlgorithm}(w_f, \zeta)$ \Comment*[r]{Bit width assignment for each layer based on Hessian matrix}
\For{$i \in [0, \ldots, L - 1]$}
{
    $\mathbb{B}^{(i)} \leftarrow \{2^{l_w^{(i)}-1}, 2^{l_w^{(i)}-2} \ldots, 2^1, 2^0\}$; \\
    \If{$\texttt{has\_{outlier}}(w^{(i)})$}  
    {
        $\mathbb{B}^{(i)} \leftarrow \{2^{l_w^{(i)}}, 2^{l_w^{(i)}-2}, \ldots, 2^1, 2^0\}$ \Comment*[r]{Bit re-weighting}
    }
}

\For{$e \in [0, \ldots, E - 1]$} 
{
    Forward propagation based on Equation \ref{eq:bsq_forward} \Comment*[r]{Re-weighted quantization}
    Compute loss $\mathcal{L}_{CE}$; \\
    Back propagation based on Equation \ref{eq:bsq_back} \Comment*[r]{Training with STE}
    Update $w^{(b)}$ with SGD optimizer;
}

\KwRet $w$ and $\mathbb{B}$

\end{algorithm}

\subsection{2PC-friendly Bit Re-weighting Algorithm}
\label{sec:winquant+}




\paragraph{Bit Re-weighting}
Based on the above Hessian-based bit width assignment, we obtain the bit width of each layer.
However, as mentioned in Section \ref{sec:motivation}, outliers introduced by Winograd transformations make quantization challenging.
Recall for OT-based linear layers, each weight is first written as $\sum_{b=0}^{l_w-1} w^{(b)} \cdot 2^b$ and then,
each bit $w^{(b)}$ is multiplied with the corresponding activations with a single OT.
This provides us with opportunities to re-weight each bit by adjusting $2^b$ to increase
the representation range without causing extra communication overhead.
We define $2^b$ as the importance for $b$-th bit and define 
$\mathbb{B}=\{2^{l_w-1}, 2^{l_w-2} \ldots, 2^1, 2^0\}$.
The OT-based computation can be re-written as $\sum_{b=0}^{l_w-1} w^{(b)} \cdot \mathbb{B}[b]$.
We first determine whether there is an outlier situation in each layer based 
on the ratio between the maximum and the standard deviation of the weights.
Then, for the layers with large outliers, we re-weight the bit by adjusting $2^b$ to $2^{b^\prime}(b^\prime>b)$
to increase the representation range flexibly. Specifically, we increase MSB by adjusting
$\mathbb{B}$ to $\{2^{l_w}, 2^{l_w-2} \ldots, 2^1, 2^0\}$ to accommodate the outliers flexibly.

\paragraph{Finetuning}
After the bit re-weighting, we further conduct quantization-aware finetuning to improve the accuracy
of the quantized networks. Since the bit importance is adjusted, we find it is more convenient to
leverage the bit-level quantization strategy \cite{yang2021bsq} to finetune our networks.
Bit-level quantization with bit representation is demonstrated in Figure \ref{fig:bit_repre}.
More specifically, the quantized weight after bit re-weighting can be formulated as
\begin{equation}
\label{eq:bsq_forward}
    \textbf{Forward:}~
    w_q = s\cdot \mathrm{Round} (\sum_{b=0}^{l_w-1} w^{(b)}\cdot \mathbb{B}[b] ) / (2^{l_w}-1),
\end{equation}
\begin{equation}
\label{eq:bsq_back}
    \textbf{Backward:}~
    \frac{\partial \mathcal{L}}{\partial w^{(b)}} = \frac{2^b}{2^{l_w}-1} \frac{\partial \mathcal{L}}{\partial w_q} = \frac{2^b}{2^{l_w}-1} \frac{\partial \mathcal{L}_{CE}(\mathcal{M}_{w_q}(x), y)}{\partial w_q},
\end{equation}
where $(x, y)$ denotes input-label pair, $s$ denotes the scaling factor, $\mathcal{L}_{CE}$ denotes cross-entropy loss for classification task, $\mathcal{M}$ denotes the model architecture.
During finetuning, $w^{(b)}$ and $s$ are trainable via STE \cite{bengio2013estimating}. 

\section{Experiments}


\paragraph{Experimental Setups}

\method~is implemented on the top of SiRNN \cite{rathee2021sirnn} in EzPC\footnote{\url{https://github.com/mpc-msri/EzPC}} library.
Following \cite{huang2022cheetah,rathee2020cryptflow2}, we use LAN modes for communication. 
For network quantization, we use quantization-aware training (QAT) for the Winograd-based networks.
We conduct experiments on different datasets and networks, i.e., MiniONN \cite{liu2017oblivious} on CIFAR-10, ResNet-32 \cite{he2016deep} on CIFAR-100, ResNet-18 \cite{he2016deep} on Tiny-ImageNet and ImageNet \cite{deng2009imagenet}.
For baselines, DeepReDuce \cite{jha2023deepreshape}, SNL \cite{cho2022selective}, SAFENet \cite{lou2020safenet}, and SENet \cite{kundu2023learning} are ReLU-optimized methods evaluated under SiRNN \cite{rathee2021sirnn}.
Please refer to Appendix \ref{setups} for details, including the network and protocol used in each baseline.

\subsection{Micro-benchmark on Efficient Protocols}

\paragraph{Convolution Protocol}
To verify the effectiveness of our proposed convolution protocol, we benchmark the convolution communication in Table \ref{tab:conv_micro_bench}.
We compare \method~with SiRNN \cite{rathee2021sirnn} and CoPriv \cite{zeng2023copriv} given different layer dimensions.
As can be observed, with W2A4,
\method~achieves 15.8$\sim$24.8$\times$ and 9.6$\sim$14.7$\times$ communication reduction, compared to SiRNN and CoPriv, respectively.

\begin{table}[!tb]
    \centering
    \begin{minipage}[t]{0.5\textwidth}
        \centering
        \caption{Micro-benchmark (MB) of convolution protocol.
        }
        \label{tab:conv_micro_bench}
        \resizebox{\linewidth}{!}{
    \begin{tabular}{cccccccc}
    \toprule
    \multicolumn{4}{c}{Dimension}  & \multirow{2}[4]{*}{SiRNN} &  \multirow{2}[4]{*}{CoPriv} & \multirow{2}[4]{*}{\method~(W2A4)} & \multirow{2}[4]{*}{\method~(W2A6)}  \\
    \cmidrule{1-4} 
     H      & W & C & K          &  &     &     &        \\ 
    \midrule
    32 & 32 & 16 & 32 &  414.1  & 247.3    &    25.75     &  30.88    \\
    16& 16& 32& 64 &   380.1 & 247.3   &  17.77     &  20.33   \\
    56& 56& 64& 64& 9336 & 5554 & 376.9  & 440.6  \\
    28& 28& 128& 128& 8598 & 5491  & 381.9  & 438.9  \\
    \bottomrule
    \end{tabular}
    }
    \end{minipage}
    \hfill
    \begin{minipage}[t]{0.42\textwidth}
        \centering
         \captionof{figure}{Micro-benchmark of residual protocol. 
         }
         \includegraphics[width=0.7\linewidth]{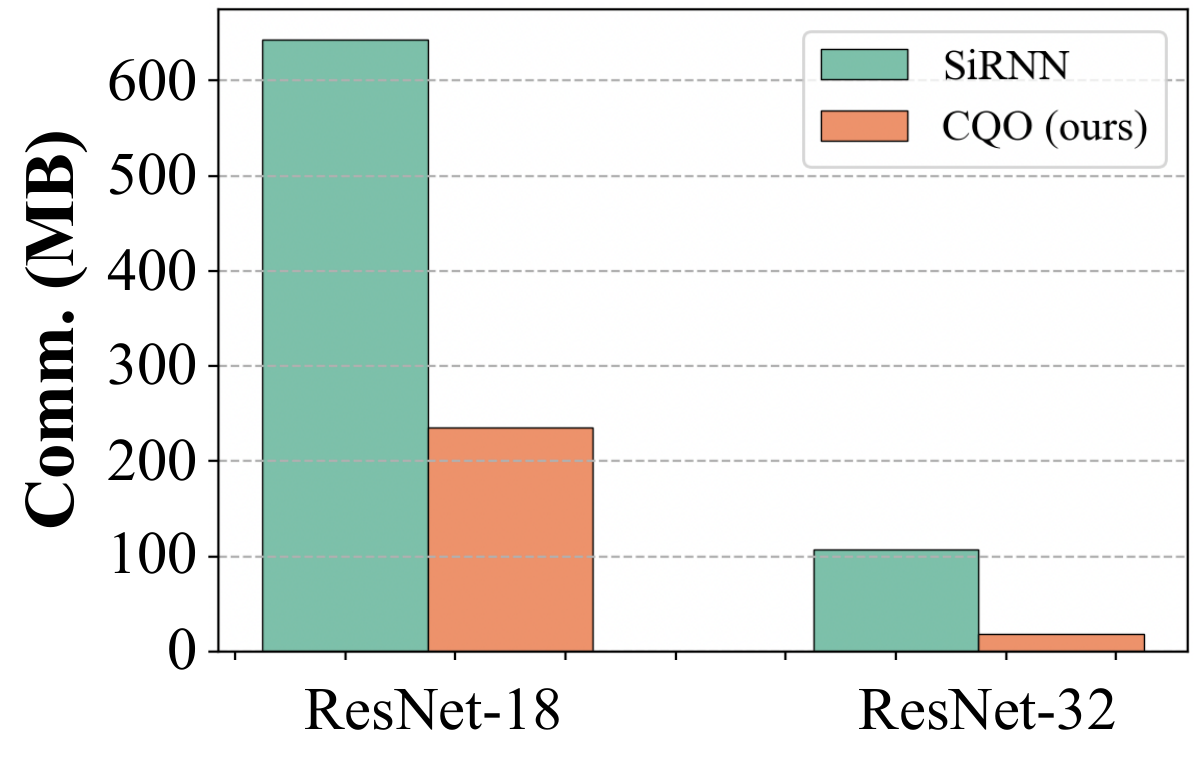}
        \label{tab:res_bench}
    \end{minipage}
\end{table}

\paragraph{Residual Protocol}
As shown in Figure \ref{tab:res_bench}, we compare our simplified residual protocol with SiRNN. 
\method~achieves 2.7$\times$ and 5.6$\times$ communication reduction on ResNet-18 (ImageNet) and ResNet-32 (CIFAR-100), respectively.


\begin{figure}[!tb]
    \centering
    \includegraphics[width=\linewidth]{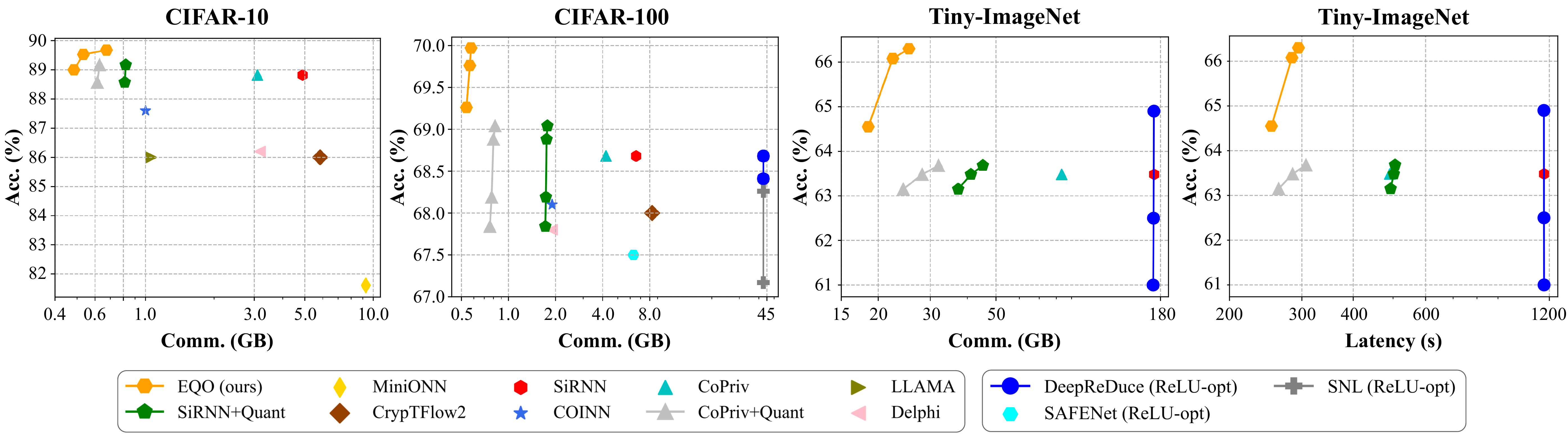}
    \caption{Comparison with prior-art methods on three datasets.
    }
    \label{fig:main_result}
\end{figure}

\subsection{Benchmark on End-to-end Inference}

\paragraph{CIFAR-10 and CIFAR-100 Evaluation}
From Figure \ref{fig:main_result}(a) and (b), we make the following observations:
\underline{1)} \method~achieves state-of-the-art (SOTA) Pareto front of the accuracy and communication. 
Specifically, on CIFAR-10, \method~outperforms SiRNN with 0.71\% higher accuracy and 9.41$\times$ communication reduction.
On CIFAR-100, \method~achieves 1.29\% higher accuracy and 11.7$\times$/6.33$\times$ communication reduction compared with SiRNN/CoPriv;
\underline{2)} compared with COINN, \method~achieves 1.4\% and 1.16\% higher accuracy as well as 2.1$\times$ and 3.6$\times$ communication reduction on CIFAR-10 and CIFAR-100, respectively;
\underline{3)} we also compare \method~with ReLU-optimized methods. 
The result shows these methods cannot effectively reduce total communication, and \method~achieves more than 80$\times$ and 15$\times$ communication reduction with even higher accuracy, compared with DeepReDuce/SNL and SAFENet, respectively;

\begin{figure}[!tb]
    \centering
    \includegraphics[width=\linewidth]{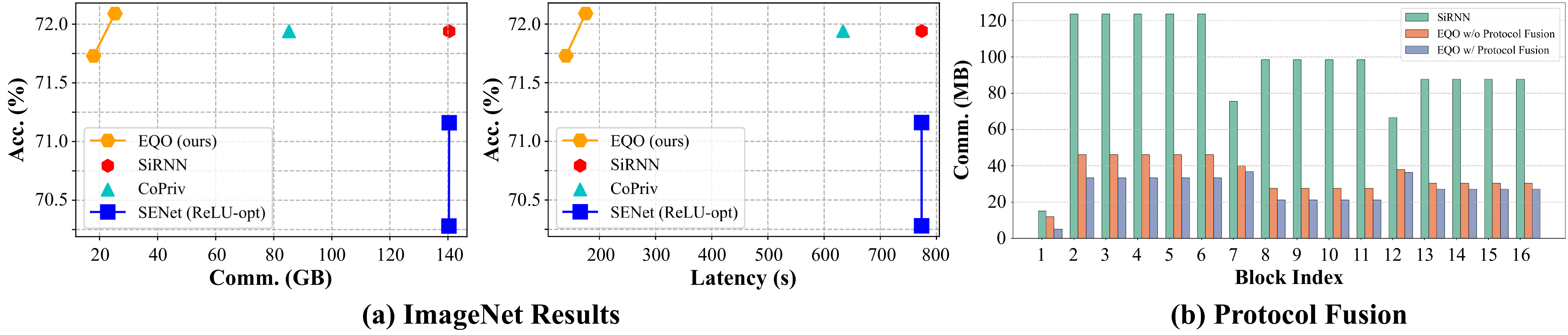}
    \caption{(a) Evaluation on ImageNet. (b) Ablation study and block-wise visualization of graph-level protocol fusion of Winograd transformation. 
    }
    \label{fig:imagenet_abl_fuse}
\end{figure}

\paragraph{Tiny-ImageNet and ImageNet Evaluation}
We compare \method~with SiRNN, CoPriv, and ReLU-optimized method DeepReDuce on Tiny-ImageNet.
As shown in Figure \ref{fig:main_result}(c), we observe that 
\underline{1)} \method~achieves 9.26$\times$ communication reduction with 1.07\% higher accuracy compared with SiRNN;
\underline{2)} compared with SiRNN equipped with mixed-precision quantization, \method~achieves 2.44$\times$ communication reduction and 0.87\% higher accuracy;
\underline{3)} compared with CoPriv, \method~achieves 4.5$\times$ communication reduction with 1.07\% higher accuracy.
We also evaluate the accuracy and efficiency of \method~on ImageNet.
As shown in Figure \ref{fig:imagenet_abl_fuse}(a), 
\method~achieves 4.88$\times$ and 2.96$\times$ communication reduction with 0.15\% higher accuracy compared with SiRNN and CoPriv, respectively.

\subsection{Ablation Study}

\begin{figure}[!tb]
    \centering
    \includegraphics[width=\linewidth]{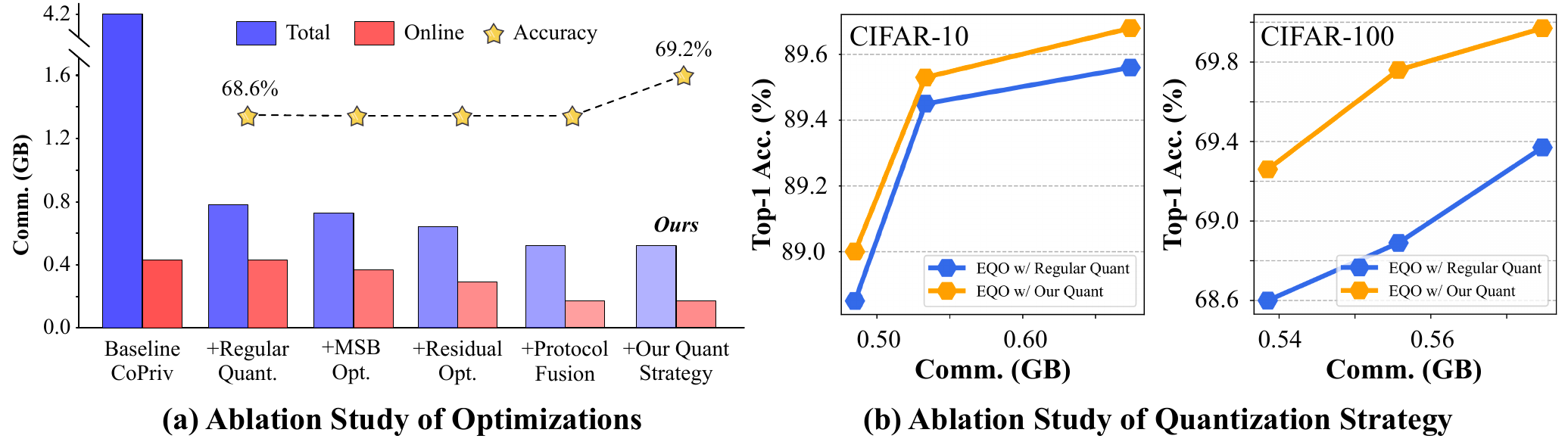}
    \caption{Ablation studies of \method.}
    \label{fig:ablation}
    \vspace{-10pt}
\end{figure}

\paragraph{Effectiveness of Quantization Strategy} 
As shown in Figure \ref{fig:ablation}(b), our proposed quantization algorithm achieves 0.1$\sim$0.2\% higher accuracy and more than 0.6\% higher accuracy on CIFAR-10 and CIFAR-100 without sacrificing efficiency, respectively, demonstrating the efficacy of \method.

\paragraph{Block-wise Visualization of Protocol Fusion}
To demonstrate the effectiveness of protocol fusion of Winograd transformation, we take W2A6 as an example to conduct the ablation experiment on ResNet-32 on CIFAR-100.
As shown in Figure \ref{fig:imagenet_abl_fuse}(b), 
protocol fusion further saves 30\% communication compared to without fusion.
We also find the communication portion of bit width conversions for Winograd transformation in the early layer is larger than in the later layers.


\paragraph{Effectiveness of Different Optimizations}
To understand how different optimizations help improve communication efficiency, we add the protocol optimizations step by step on ResNet-32, and present the results in Figure \ref{fig:ablation}(a).
As observed from the results, we find
that
\underline{1)} low-precision Winograd quantization benefits the total communication efficiency most. However, there is no benefit to online communication due to the extra bit width conversions for Winograd transformation even though it is low precision;
\underline{2)} simplified residual protocol, protocol fusion, and MSB-known optimization consistently reduce the online and total communication;
\underline{3)} although naively combining Winograd convolution with quantized private inference enlarges the online communication, our optimizations finally achieve 8.1$\times$ and 2.5$\times$ total and online communication reduction compared to CoPriv;
\underline{4)} \method~with bit re-weighting achieves higher accuracy without sacrificing efficiency.
The findings indicate that all of our optimizations are indispensable for 2PC-based inference efficiency.

\section{Conclusion}

In this work, we propose \method~, a communication-efficient 2PC-based framework with Winograd and mixed-precision quantization.
We observe naively combining quantization and Winograd convolution is sub-optimal.
Hence, at the protocol level, we propose a series of optimizations for the 2PC inference graph to minimize the communication.
At the network level, we develop a sensitivy-based mixed-precision quantization algorithm and a 2PC-friendly bit re-weighting algorithm to accommodate weight outliers without increasing bit widths.
With extensive experiments, \method~consistently reduces the communication without compromising the accuracy compared with the prior-art 2PC frameworks and network optimization methods.



%
%

\bibliographystyle{splncs04}
\bibliography{main}

\clearpage

\appendix


\section{Threat Model}
\label{supp:threat}

\method~is a 2PC-based private inference framework that involves a client and a server.
The server holds the private model weights and the client holds the private input data.
Following \cite{rathee2021sirnn,rathee2020cryptflow2,mohassel2017secureml,mishra2020delphi,cho2022selective,jha2021deepreduce,kundu2023learning},
we assume the DNN architecture (including the number of layers and the operator type, shape, 
and bit widths) are known by two parties.
At the end of the protocol execution, the client learns the inference result and the two parties know nothing else
about each other's input.
Following prior work, we assume the server and client are semi-honest adversaries. 
Specifically, both parties follow the protocol specifications but also attempt to learn more from the information than allowed.
We assume no trusted third party exists so the helper data needs to be generated by the client and server \cite{rathee2020cryptflow2,rathee2021sirnn,mishra2020delphi}.

\section{Notations and Underlying Protocols}
\label{supp:notation}
Notations used in this paper is shown in Table \ref{tab:supp:notation}.
We also show the detailed communication complexity with and without MSB-known optimization of the underlying protocols in Table \ref{tab:prot_comm}.

\begin{table}[!tb]
    \centering
    \caption{Notations and the corresponding descriptions used in this paper. 
    }
    \label{tab:supp:notation}
    \resizebox{\linewidth}{!}{
    \begin{tabular}{c|c}
    \toprule
    Notation     & Description \\
    \hline
    $\lambda$ & Security parameter \\
    $\gg $ & Shift right\\
    $l, s $ & Bit width, scale \\
    $l_w, l_a, l_{acc}, l_{res}, l_{add}$ & \tabincell{c}{Bit width of weight, activation, accumulation, residual, and addition} \\
    $l_{feature\_ext}, l_{out\_ext}$  & \tabincell{c}{Bit width of Winograd feature and output transformation} \\
    $x^{(l)}, \langle x \rangle^{(l)}$ & An $l$-bit integer $x$ and $l$-bit secret shares \\
    $H, W, C, K$  & Height and width of output feature, and number of input and output channel \\
    $A, B, G$  & Winograd transformation matrices \\
    $m, r, T$  & Size of output tile and convolution weight, and number of tiles \\
    W$m$A$n$ & $m$-bit weight and $n$-bit activation   \\
    \bottomrule
    \end{tabular}
    }
\end{table}

\section{Supplementary Preliminaries}
\label{supp:preli}

\subsection{Arithmetic Secret Sharing}
\label{supp:arss}
Arithmetic secret sharing (ArSS) is a fundamental cryptographic scheme used in our framework.
Specifically, an $l$-bit value $x$ is additively shared in the integer ring $\mathbb{Z}_{2^l}$ as the sum of two values,
e.g., $\langle x\rangle_s^{(l)}$ held by the server and $\langle x\rangle_c^{(l)}$ held by the client. 
$x$ can be reconstructed as $\langle x\rangle_s^{(l)} + \langle x\rangle_c^{(l)} \mod 2^l$.
With ArSS, additions can be conducted locally by the server and client while multiplication requires helper data,
which are independent of the secret shares and are generated through communication during the pre-processing stage \cite{mishra2020delphi,rathee2020cryptflow2,rathee2021sirnn}.

\subsection{Winograd Convolution}
Winograd convolution \cite{lavin2016fast} is an efficient convolution algorithm that minimizes the number of multiplications when computing a convolution.
The 2D Winograd transformation $F(m \times m, r \times r)$, where the sizes of the output tile, weight, and input tile are
$m \times m$, $r \times r$, and $n \times n$, respectively, with $n = m + r -1$, can be formulated as follows:
\begin{equation*}
    \label{eq:winograd}
    Y = W \circledast X = A^\top [(G W G^\top) \odot (B^\top X B)]A,
\end{equation*}
where $\circledast$ denotes a regular convolution and $\odot$ denotes element-wise matrix multiplication (EWMM). 
$A$, $B$, and $G$ are transformation matrices
that are independent of the weight $W$ and activation $X$ and can be computed based on $m$ and $r$
\cite{lavin2016fast,alam2022winograd}.
When computing the Winograd convolution, $W$ and $X$ are first transformed to the Winograd domain before computing EWMM.
The number of multiplications can be reduced from $m^2r^2$ to $(m+r-1)^2$ at the cost of many more additions
introduced by the transformations of $B^\top X B$ and $A^\top [\cdot] A$.
\cite{zeng2023copriv} further formulates the EWMM into the form of general matrix multiplication for better
communication efficiency.
As a consequence, the communication complexity is reduced from $O(\lambda CKT(m+r-1)^2)$ to $O((m+r-1)^2CT(\lambda+K))$,
where $C, K$, and $T$ denote the number of input channels, output channels, and the number of tiles, respectively.


\subsection{Network Quantization}
\label{supp:quant}


Quantization converts a floating-point number into integers for a better training and inference efficiency \cite{krishnamoorthi2018quantizing}.
Specifically, a floating point number $x_f$ can be approximated by an $l_x$-bit integer $x_q$ and a scale $s_x$ 
through quantization as $x_q / s_x$,
where
\begin{equation*}
    x_q = \max(-2^{l_x-1}, \min(2^{l_x-1}-1, \mathrm{round}(s_x x_f))).
\end{equation*}

The multiplication of two floating point numbers 
$x_f$ and $w_f$, denoted as $y_f$, can be approximately computed as $x_q w_q / (s_w s_x)$,
which is a quantized number with $(l_x + l_w)$-bit and $s_w s_x$ scale.
Then, $y_f$ usually needs to be re-quantized to $y_q$ with $l_y$-bit and $s_y$ scale as follows:
\begin{equation*}
    y_q = \max(-2^{b_y-1}, \min(2^{b_y-1}-1, \mathrm{round}(\frac{s_y}{s_w s_x} w_q x_q))).
\end{equation*}

For the addition, e.g., residual connection of two quantized numbers $x_q$ and $y_q$, directly adding them together leads to incorrect results. Instead, the scales and the bit widths of $x_q$ and $y_q$ need
to be aligned first. 



\begin{table}[!tb]
    \caption{Communication complexity of underlying protocols.}
    \label{tab:prot_comm}
    \centering
    \resizebox{\linewidth}{!}{
    \begin{tabular}{c|c|c}
    \toprule
     Protocol & Communication w/o MSB Optimization  & Communication w/ MSB Optimization   \\ 
    \hline
    $\Pi_{\mathrm{Ext}}^{l_1,l_2}(\langle x\rangle^{(l_1)})$ &  $O(\lambda (l_1+1)+13l_1+l_2)$  &   $O(2\lambda -l_1+l_2+2)$ \\ 
    $\Pi_{\mathrm{Trunc}}^{l_1,l_2}(\langle x\rangle^{(l_1)})$ &  $O(\lambda (l_1+3)+15l_1+l_2+20)$  &  $O(3\lambda +l_1+l_2+20)$  \\ 
    $\Pi_{\mathrm{TR}}^{l_1,l_2}(\langle x\rangle^{(l_1)})$ &  $O(\lambda (l_2+1)+13l_2+l_1)$ &  $O(\lambda+2)$ \\ 
    \bottomrule
    \end{tabular}
    }
\end{table}

\section{Related Works}
\label{related}

\noindent \textbf{Efficient Private Inference}
In recent years, there has been an increasing amount of literature on efficient private inference, including protocol optimization \cite{demmler2015aby,mohassel2018aby3,rathee2020cryptflow2,rathee2021sirnn,knott2021crypten,rathee2022secfloat,mohassel2017secureml,hao2022iron}, network optimization \cite{cho2022selective,jha2021deepreduce,kundu2023learning,li2022mpcformer,zeng2023mpcvit,lou2020safenet} and co-optimization \cite{mishra2020delphi,hussain2021coinn,zeng2023copriv,ran2022cryptogcn,peng2023lingcn}.
These works are designed for convolutional neural networks (CNNs), Transformer-based models and graph neural networks (GNNs).
In this paper, we mainly focus on efficient quantized private inference.
XONN \cite{riazi2019xonn} combines binary neural networks with Yao's garbled circuits (GC) to replace the costly multiplications with XNOR operations.
As an extension of XONN, \cite{samragh2021application} proposes a hybrid approach where the 2PC protocol is customized to each layer.
CrypTFlow2 \cite{rathee2020cryptflow2} supports efficient private inference with uniform bit widths of both weight and activation.
SiRNN \cite{rathee2021sirnn} proposes a series of protocols to support bit extension and truncation, which enable mixed-precision private inference for recurrent neural networks (RNNs).
ABNN2 \cite{shen2022abnn2} utilizes the advantages of quantized network and realizes arbitrary-bitwidth quantized private inference by proposing efficient matrix multiplication protocol based on 1-out-of-N OT extension.
COINN \cite{hussain2021coinn} proposes low-bit quantization for efficient ciphertext computation, and has 1-bit truncation error.

\noindent \textbf{Winograd Algorithm}
Many research efforts have been made to Winograd algorithm in hardware domain \cite{wang2020winonn,kala2019high,jia2020enabling,jorda2022cuconv,yan2020optimizing}.
\cite{lavin2016fast} first applies Winograd algorithm to CNNs, and shows 2.25$\sim$4$\times$ efficiency improvement.
\cite{fernandez2020searching} applies Winograd convolution to 8-bit quantized networks, and adds transformation matrices to the set of learnable parameters. \cite{fernandez2020searching} also proposes a Winograd-aware neural architecture search (NAS) algorithm to select among different tile size, achieving a better accuracy-latency trade-off.
\cite{liu2018efficient} exploits the sparsity of Winograd-based CNNs, and propose two modifications to avoid the loss of sparsity.
For Winograd quantization, \cite{chikin2022channel} proposes to balance the ranges of input channels on the forward pass by scaling the input tensor using balancing coefficients.
To alleviate the numerical issues of using larger tiles, \cite{andri2022going} proposes a tap-wise quantization method.
\cite{ganesan2022efficient} integrates Winograd convolution in CrypTFlow2 \cite{rathee2020cryptflow2}, but only uses 60-bit fixed-point number, leading to significant communication overhead.
Therefore, quantized Winograd-based networks for privacy inference have not been well studied.

\section{Details of Experimental Setups}
\label{setups}

\noindent \textbf{Private Inference}
\method~is implemented based on the framework of SiRNN \cite{rathee2021sirnn} in the EzPC\footnote{\url{https://github.com/mpc-msri/EzPC}} library for private network inference.
Specifically, our quantized Winograd convolution is implemented in C++ with Eigen and Armadillo matrix calculation library \cite{sanderson2016armadillo}.
Following \cite{huang2022cheetah,rathee2020cryptflow2}, we use WAN and LAN modes for 2PC communication. 
Specifically, the bandwidth between is 377 MBps and 40 MBps,
and the echo latency is 0.3ms and 80ms in LAN and WAN mode, respectively.
In our experiments, we evaluate the inference efficiency on the
Intel Xeon Gold 5220R CPU @ 2.20GHz.

\begin{figure}[!tb]
    \centering
    \includegraphics[width=\linewidth]{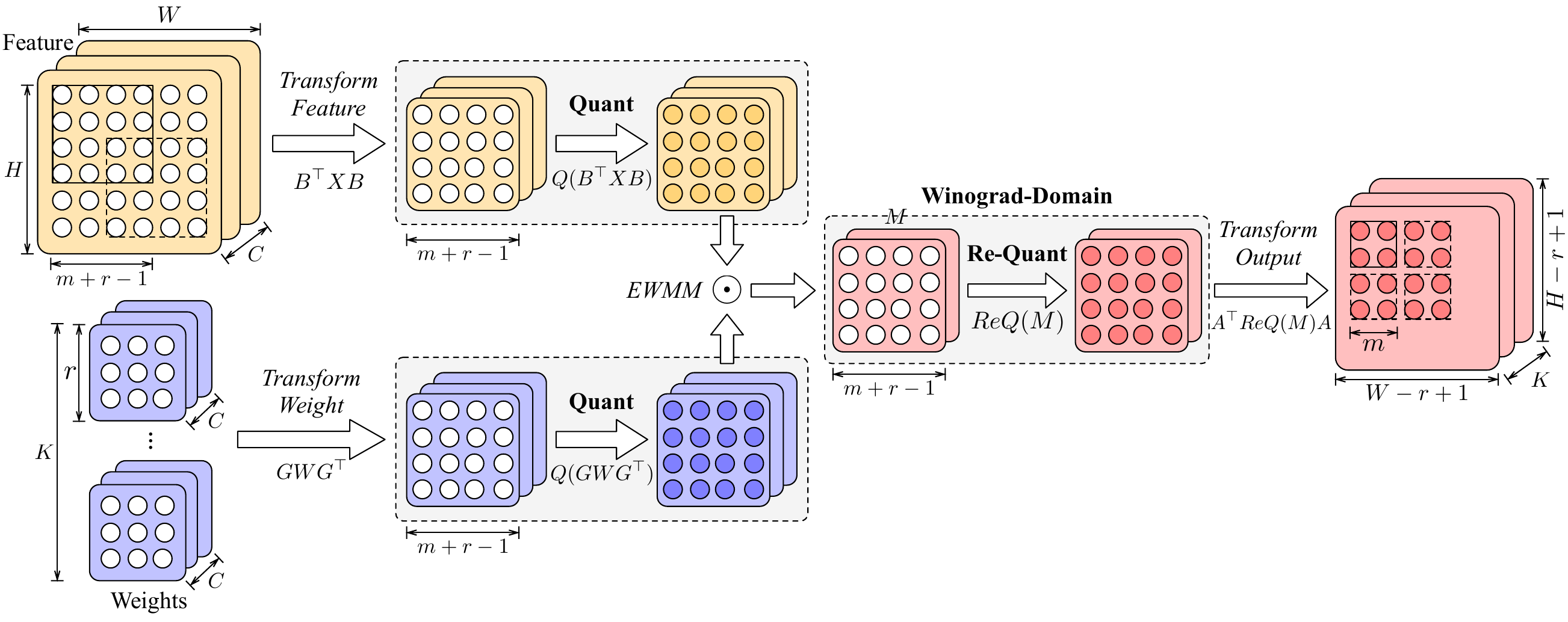}
    \caption{Overview of quantization procedure of Winograd convolution.}
    \label{fig:winograd_qat}
\end{figure}

\noindent \textbf{Network Quantization}
To improve the performance of quantized networks, we adopt quantization-aware training with PyTorch framework for the Winograd-based networks.
During training, we fix the transformation matrices $A, B$ while $G$ is set to be trainable since $GWG^\top$ is computed and quantized before inference (processed offline).
For each network, we fix the bit width of the first and last layer to 8-bit. For MiniONN, we fix the bit width of activation to 4-bit while for ResNets, we fix it to 6-bit.
As we have introduced, we always use high precision, i.e., 8 bits, for the residual for better accuracy \cite{wu2018mixed,yang2021fracbits}. The benefit
of high-precision residual is evaluated in Appendix \ref{app:high_precision_res}.
We illustrate the quantization procedure for Winograd convolution with $F(2\times 2, 3\times 3)$ transformation as shown in Figure \ref{fig:winograd_qat}.
Following \cite{li2021lowino,fernandez2020searching,li2020lance,krishnamoorthi2018quantizing}, 
we use fake quantization to both the activation and weights and then re-quantize the activation back after GEMM.

\noindent \textbf{Baselines}
We describe the detailed information of each baseline we use in Table \ref{tab:baselines}, including their networks, datasets, and protocols.

\begin{table}[!tb]
    \centering
    \caption{Networks, datasets, and protocols used in baselines.}
    \label{tab:baselines}
    \resizebox{0.85\linewidth}{!}{
    \begin{tabular}{ccc}
    \toprule
    Method     &  Network (Dataset)  &   Protocol  \\
    \hline
       & Protocol Frameworks  &    \\
    \hline
    CrypTFlow2 \cite{rathee2020cryptflow2}   & \tabincell{c}{MiniONN (CIFAR-10), ResNet-32 (CIFAR-100), \\ ResNet-18 (Tiny-ImageNet, ImageNet)} &  OT  \\
    SiRNN \cite{rathee2021sirnn}   & \tabincell{c}{MiniONN (CIFAR-10), ResNet-32 (CIFAR-100), \\ ResNet-18 (Tiny-ImageNet, ImageNet)} &  OT  \\
    CoPriv \cite{zeng2023copriv}   &  \tabincell{c}{MiniONN (CIFAR-10), ResNet-32 (CIFAR-100), \\ ResNet-18 (Tiny-ImageNet, ImageNet)}   & OT \\
    COINN \cite{hussain2021coinn}  &  MiniONN (CIFAR-10), ResNet-32 (CIFAR-100)  & OT   \\
    MiniONN \cite{liu2017oblivious}  &  MiniONN (CIFAR-10) & OT  \\
    \hline
       &  ReLU-optimized Methods  &   \\
    \hline
    DeepReDuce \cite{jha2021deepreduce}    &  ResNet-18 (CIFAR-100, Tiny-ImageNet) &   OT   \\
    SNL \cite{cho2022selective}  &    ResNet-18 (CIFAR-100) &  OT \\
    SAFENet \cite{lou2020safenet}          &   ResNet-32 (CIFAR-100)     &  OT  \\
    SENet \cite{kundu2023learning}  & ResNet-18 (ImageNet)  &  OT  \\
    \bottomrule
    \end{tabular}
    }
\end{table}

\section{Benefit of High-precision Residual} 
\label{app:high_precision_res}
\begin{figure}
    \centering
    \includegraphics[width=0.6\linewidth]{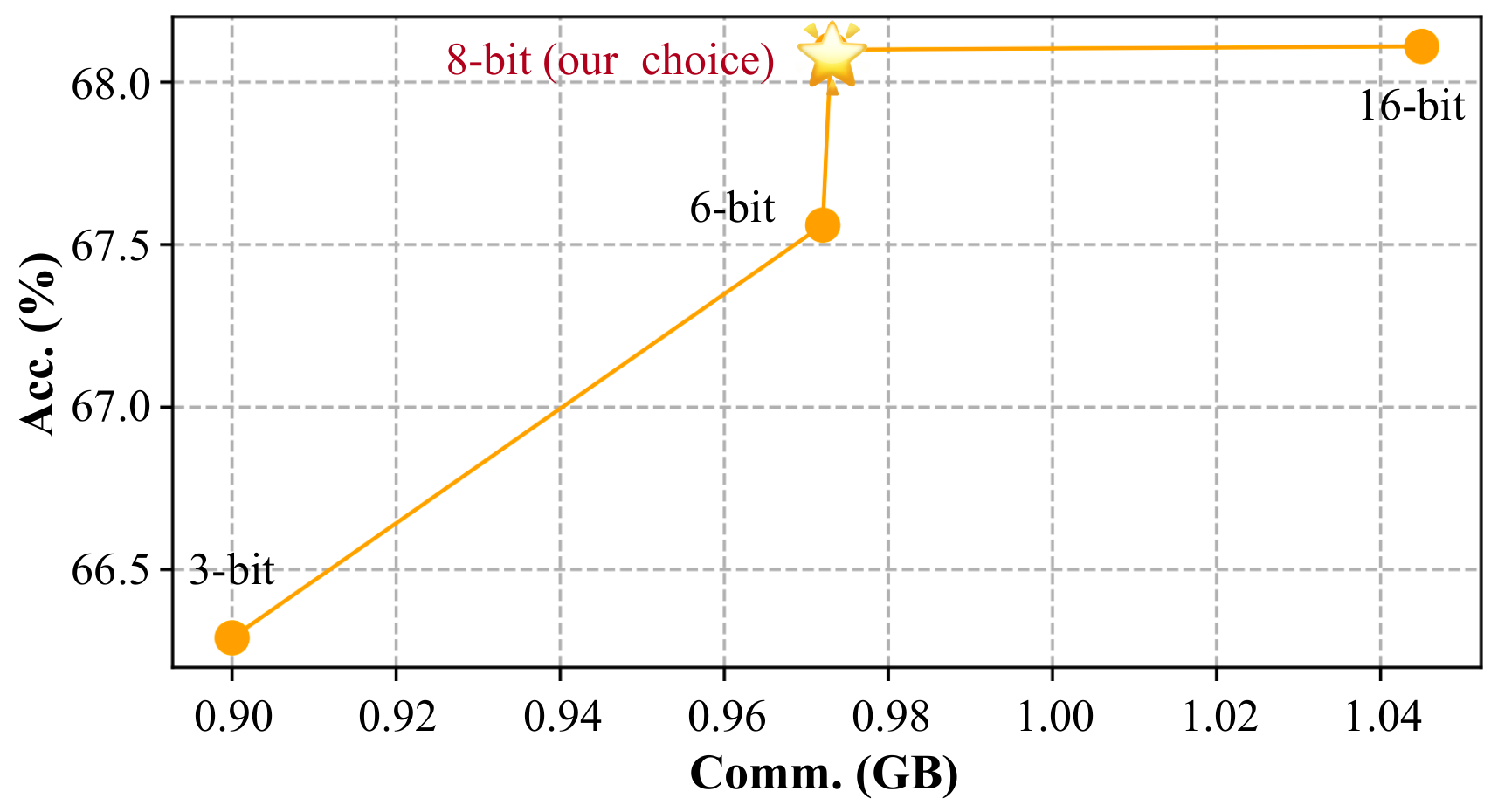}
    \caption{The impact of residual bit width on accuracy and communication for ResNet-32 on CIFAR-100.}
    \label{fig:res_bw}
\end{figure}

From Figure \ref{fig:res_bw}, we fix the bit width of activation and weight to 3-bit and 4-bit, we observe that the accuracy of ResNet-32 significantly improves by 1.8\% and the communication only increases slightly when we increase the residual bit width from 3-bit to 8-bit. 
And the benefit of 16-bit residual is not significant.
Hence, we propose to use 8-bit high-precision residual to train our quantized networks for a better performance.


\section{Online and Total Communication Comparison with ReLU-optimized Methods}
\label{supp:online}

In Table \ref{tab:online}, we separately compare the online and total communication with ReLU-optimized methods.
The result shows that \method~focuses reducing the dominant convolution communication, such that achieves significantly lower total communication with even higher accuracy.
In terms of online communication, although \method~does not focus on directly removing online components, e.g., ReLUs, \method~still achieves low online communication with our proposed graph-level optimizations.

\begin{table}[!tb]
    \centering
    \caption{Comparison of online and total communication on the CIFAR-100 dataset.}
    \label{tab:online}
    \resizebox{0.7\linewidth}{!}{
    \begin{tabular}{cccc}
    \toprule
    Method    &  Online (GB) & Total (GB) & Acc. (\%) \\
    \hline
    DELPHI \cite{mishra2020delphi}     &  0.19   & 1.90 & 67.8  \\
    DeepReDuce \cite{jha2021deepreduce} &  0.15   & 42.6 &  68.4    \\
    \method    &  0.17   & 0.47  &  69.2 \\
    \bottomrule
    \end{tabular}
    }
\end{table}

\section{Analysis of the Extension Communication Complexity of Winograd Transformation}
\label{app:extension}

As mentioned in Section \ref{supp:notation}, the communication complexity per element mainly scales with the initial bit width $l_1$.
Also, for a given matrix with the dimension of $d_1\times d_1\times d_3$, the number of extensions needed would be $d_1\times d_1\times d_3$.
Therefore, the total communication complexity of extension for a matrix becomes $O(d_1d_2d_3(\lambda (l_1+1)+13l_1+l_2))$.
In Figure \ref{fig:protocol}, the extension communication of feature extension (block \ding{173}) is $O(HWC(\lambda (l_1+1)+13l_1+l_2))$ and the extension communication of output extension (block \ding{175}) is $O(KT(m+r-1)^2(\lambda (l_1+1)+13l_1+l_2))$, which is more expensive for communication.

\section{Formal Description of Protocol Fusion}
\label{supp:fusion}

\begin{proposition}
\label{prop:decomp}
    For a given $\langle x\rangle^{(l_1)}$, 
    $\Pi_{\mathrm{Trunc}}^{l_1, l_2}(\langle x\rangle^{(l_1)})$ can be decomposed into $\Pi_{\mathrm{TR}}^{l_1, l_2}$ followed by $\Pi_{\mathrm{Ext}}^{l_1-l_2, l_1}$ as
    \begin{align*}
        \Pi_{\mathrm{Trunc}}^{l_1, l_2}(\langle x\rangle^{(l_1)}) & = \Pi_{\mathrm{Ext}}^{l_1 - l_2, l_1}( \Pi_{\mathrm{TR}}^{l_1, l_2}(\langle x\rangle^{(l_1)})).
    \end{align*}
    The decomposition does not change communication.
\end{proposition}

\begin{proposition}
\label{prop:ext}
    If a given $\langle x\rangle^{(l_1)}$ is extended to $l_2$-bit first and then extended to $l_3$-bit. Then, the two neighboring extensions can be fused together as
    \begin{align*}
        \Pi_{\mathrm{Ext}}^{l_2, l_3}(\Pi_{\mathrm{Ext}}^{l_1, l_2}(\langle x\rangle^{(l_1)})) & = \Pi_{\mathrm{Ext}}^{l_1, l_3}(\langle x\rangle^{(l_1)}).
    \end{align*}
    Extension fusion reduces communication from $O(\lambda(l_1+l_2+2))$ to $O(\lambda(l_1+1))$. 
\end{proposition}

\begin{proposition}
    For a given $\langle x\rangle^{(l_1)}$,
    when re-quantization ends up with truncation, and is followed by an extension, the protocol can be first decomposed and then fused as
    \begin{align*}
        \Pi_{\mathrm{Ext}}^{l_1, l_3}( \Pi_{\mathrm{Trunc}}^{l_1, l_2}(\langle x\rangle^{(l_1)}) ) = \Pi_{\mathrm{Ext}}^{l_1, l_3}( \Pi_{\mathrm{Ext}}^{l_1 - l_2, l_1}( \Pi_{\mathrm{TR}}^{l_1, l_2}(\langle x\rangle^{(l_1)}))) 
        = \Pi_{\mathrm{Ext}}^{l_1 - l_2, l_3}( \Pi_{\mathrm{TR}}^{l_1, l_2}(\langle x\rangle^{(l_1)})).
    \end{align*}
    Combining Proposition \ref{prop:decomp} and \ref{prop:ext}, this fusion reduces communication by around 2$\times$ from $O(\lambda(2l_1+4))$ to $O(\lambda(l_1+2))$.
\end{proposition}

\end{document}